\documentclass[12pt]{iopart}
\usepackage{graphicx,url,longtable,rotating,color,units,wasysym,subfig,epsfig,amssymb,amstext}
\usepackage[normalem]{ulem}
\usepackage[square, comma, sort&compress]{natbib}

\begin{document}

\title[Identification of noise artifacts in searches for long GW transients]{Identification of noise artifacts in searches for long-duration gravitational-wave transients}

\author{Tanner~Prestegard$^1$, 
Eric~Thrane$^1$, 
Nelson~L.~Christensen$^2$, 
Michael~W.~Coughlin$^2$, 
Ben~Hubbert$^2$, 
Shivaraj~Kandhasamy$^1$,
Evan~MacAyeal$^2$, 
Vuk~Mandic$^1$
}

\address{$^1$ School of Physics and Astronomy,
University of Minnesota, Minneapolis, MN 55455, USA}
\address{$^2$ Physics and Astronomy, Carleton College,
Northfield, MN 55057, USA}
\ead{ethrane@physics.umn.edu}

\begin{abstract}
  We present an algorithm for the identification of transient noise artifacts (glitches) in cross-correlation searches for long gravitational-wave transients lasting seconds to weeks.
  The algorithm utilizes the auto-power in each detector as a discriminator between well-behaved stationary noise (possibly including a gravitational-wave signal) and non-stationary noise transients.
  We test the algorithm with both Monte Carlo noise and time-shifted data from the LIGO S5 science run and find that it removes a significant fraction of glitches while keeping the vast majority ($99.6\%$) of the data.
We show that this cleaned data can be used to observe GW signals at a significantly lower amplitude than can otherwise be achieved.
  Using an accretion disk instability signal model, we estimate that the algorithm is accidentally triggered at a rate of less than $10^{-5}\%$ by realistic signals, and less than $3\%$ even for exceptionally loud signals.
  We conclude that the algorithm is a safe and ef{\kern0pt}fective method for cleaning the cross-correlation data used in searches for long gravitational-wave transients.
\end{abstract}

\submitto{\CQG}
\maketitle

\section{Introduction}\label{intro}
Our aim is to detect long-lasting gravitational-wave (GW) transients (lasting seconds to weeks) in the presence of ``glitches'': non-stationary noise artifacts that contaminate the otherwise approximately Gaussian strain noise in GW interferometers.
We focus our attention on the cross-correlation method of~\cite{stamp}, though, it may be possible to extend this formalism to other search algorithms as well---a topic of ongoing research.
Possible sources of long GW transients include convection in proto-neutron stars~\cite{ott:09,dessart:06pns,mueller:04,keil:96,miralles:00,miralles:04}, rotational instabilities associated with nascent neutron stars~\cite{ott:09,corsi,piro:11,ott:07prl,ou:04,scheidegger:10b,lai:95}, instabilities in the disks of accreting systems~\cite{piro,vanPutten,vanputten:01,vanputten:08}, neutron star glitches~\cite{pglitch_strain}, soft gamma repeaters / anomalous X-ray binaries~\cite{glampedakis,samuelsson,levin,sotani,horvath,deFreitas,ioka} and dynamically formed black hole binaries~\cite{birjoo,janna,oleary,kocsis}.

Glitches can arise from environmental contamination such as mechanical vibrations, electromagnetic disturbances, circuit breaker trips, power shorts and asymmetric photodiode response~\cite{lsc_glitch}.
While some glitches can be identified and removed by comparing GW strain channels with environmental and sub-system monitoring channels, many remain after the first stages of data cleaning (see, e.g.,~\cite{smith_glitch,lsc_glitch,ajith_glitch,stamp_pem,isogai,ballinger,christensen,slutsky}).
These remaining glitches require special attention for two reasons.
First, a high glitch rate can diminish the sensitivity of a search by raising the threshold required for an event to be statistically significant\footnote{The astute reader may wonder how the present concern about glitches should be squared with the finding in~\cite{stamp} that the $\text{SNR}$ distributions for time-shifted and Monte Carlo ``are in qualitative agreement.''
Do we really need to worry about glitches in searches for long GW transients?
The answer is yes.
The results presented in~\cite{stamp} compared the standard deviation and approximate shape of distributions of {\em pixel} $\text{SNR}$ for Monte Carlo and time-shift data.
While this comparison showed that the distributions are similar, our present analysis focuses on the high-$\text{SNR}$ tail of the distribution of {\em clusters} of pixels.
Since glitches tend to produce clusters of pixels of non-Gaussian noise, their importance is magnified when we study the distribution of cluster $\text{SNR}$.
}.
Indeed, below we shall show a realistic example wherein the required signal power for a $p=0.1\%$ false alarm probability event drops two-fold when we use our algorithm to remove glitchy segments from GW data.
This level of improvement is not achievable with the application of existing data-quality flags.
Second, robust glitch identification methods can improve our confidence in a GW candidate if it does not resemble non-stationary noise.

We describe an algorithm to check the consistency of the auto-power from two terrestrial GW detectors to identify glitches in searches using the cross-power statistic described in~\cite{stamp}.
(Throughout, we use the expressions ``auto-power'' and ``cross-power'' instead of ``power spectrum,'' which can refer to either.)
%
%
We demonstrate the ability of the algorithm to improve the sensitivity of targeted searches by cleaning real interferometer data to a level approaching optimally well-behaved Gaussian noise.

This work builds on~\cite{stamp_pem}, which described how environmental monitoring channels can be used to identify {\em long-lasting noise} transients.
However, it dif{\kern0pt}fers because first, {\em we utilize only GW strain channels}, and second, because we are interested in recovering {\em long-lasting GW signals} in the presence of what are sometimes very {\em short bursts of noise}.
It also dif{\kern0pt}fers from~\cite{cannon_strings,hild_glitch} and other consistency-check algorithms that the authors are aware of because we are not checking the consistency of GW triggers, but rather we are checking the consistency of {\em data segments}---many of which will together constitute a GW trigger.
This is born of necessity from our focus on long transients.
We shall see in Section~\ref{algorithm} that by flagging individual segments as glitchy, we are able in principle to observe a GW event temporarily disturbed by non-stationary noise.

To illustrate our glitch identification algorithm, we use Monte Carlo and time-shifted data from the $\unit[4]{km}$ LIGO H1 and L1 interferometers~\cite{iligo} in Hanford, WA and Livingston, LA, respectively.
Time-shifting one strain time series with respect to another by an amount greater than the GW travel time between interferometers removes astrophysical signals while preserving non-Gaussian noise artifacts that are otherwise dif{\kern0pt}ficult to simulate.
Our Monte Carlo assumes Gaussian noise with an initial LIGO design sensitivity, and our time-shifted data are from the Nov.~5, 2005 - Sep.~30, 2007 S5 science run (see, e.g.,~\cite{stoch-S5,crab-S5}).
During S5, the LIGO interferometers achieved a strain sensitivity of $\approx\unit[3\times10^{-23}]{Hz^{-1/2}}$ in the most sensitive band around $\unit[100\sim200]{Hz}$.
We utilize a few days of accumulated data from GPS=$816065659 - 819039020$.
By comparing how the glitch identification algorithm performs for Monte Carlo and time-shifted results, we can measure how close we can get to ideal Gaussian noise by cleaning non-Gaussian noise.
While we use the LIGO H1 and L1 detectors for illustrative purposes, we expect that these techniques can be extended to additional pairs of detectors including interferometers such as Virgo~\cite{Virgo,Virgo2,Virgo3,Virgo_url}, LCGT~\cite{LCGT,LCGT_url} and GEO~\cite{GEO,GEO2,GEO3,GEO4}.

The outline for the rest of this paper is as follows.
In Section~\ref{formalism} we summarize the cross power-based analysis framework from~\cite{stamp}.
In Section~\ref{Xi} we develop an autopower dif{\kern0pt}ference statistic that can be used to evaluate whether the autopower in a pair of detectors is consistent with noise plus a GW signal.
We analyze the behavior of this statistic for stationary noise, signals, and glitches.
In Section~\ref{algorithm} we present a glitch identification algorithm based on the autopower dif{\kern0pt}ference statistic and demonstrate its ability to clean time-shifted LIGO data.
In Section~\ref{toy} we introduce an accretion disk instability waveform, which we use in  Section~\ref{safety} to investigate the safeness of our algorithm, i.e., the probability that it falsely identifies a signal as glitch-like.
In Section~\ref{DQ} we investigate the complementarity of our algorithm to data quality flags based on instrumental and environmental noise artifacts.
Section~\ref{conclusions} contains concluding remarks.

\section{Formalism}\label{formalism}
Our starting point is~\cite{stamp}, which is described in greater detail in \ref{app}.
We use the cross-correlation of two or more spatially separated interferometers to construct a statistic $\hat{Y}(t;f)$, which is an unbiased estimator for the GW power $H(t;f)$ between times $t$ and $t+\delta t$ in some frequency bin between $f$ and $f+\delta f$.
$H(t;f)$ is defined in terms of the  GW field Fourier coef{\kern0pt}ficients, $\tilde{h}_A(f)$ (see \ref{app}):
\begin{equation}\label{eq:HAAdef_body}
  H(t;f) = \text{Tr}\left[H_{AA'}(t;f)\right] =
  \text{Tr}\left[
  \frac{2}{\cal N} \langle \tilde{h}_A(t;f) \tilde{h}_{A'}(t;f) \rangle
  \right]
\end{equation}
Here the brackets $\langle...\rangle$ denote the expectation value of the enclosed quantity.
The semicolon emphasizes that $t$ refers to the beginning of a data segment of length $\delta t$ and not to the many sampling times associated with each segment.
It is important that the noise in the two interferometers is uncorrelated, which is easily achieved for spatially separated interferometers.

The set of $\hat{Y}(t;f)$ can be represented as an $ft$-map (spectrogram).
The same is true of $\hat\sigma(t;f)$, an estimator for the uncertainty associated with $\hat{Y}(t;f)$.
GW candidates are identified as clusters of high $\text{SNR}\equiv\hat{Y}/\hat\sigma$ pixels~\cite{stamp}.
The significance of a cluster $\Gamma$ can be estimated by calculating the total SNR for the entire cluster, denoted $\text{SNR}(\Gamma)$, and comparing it to the distribution of $\text{SNR}(\Gamma)$ obtained with time-shifted data~\cite{stamp}.

For suf{\kern0pt}ficiently long signals, the ef{\kern0pt}fect of non-stationary noise is averaged away and $\text{SNR}(\Gamma)$ becomes Gaussian distributed by the central limit theorem.
This limiting case is the stochastic radiometer---a technique for mapping the GW sky with two or more spatially separated interferometers~\cite{radiometer,stefan,sph_results}.
Here, however, we study (relatively) shorter time scales where glitches play a role in our ability to determine the significance of an event.
The question we aim to investigate in the rest of the paper is: how can we discriminate between large values of $\text{SNR}(\Gamma)$ due to a GW signal and large values due to glitches?

Our glitch identification algorithm will utilize cross-power $\hat{C}_{IJ}(t;f)$ and auto-power $\hat{P}_I(t;f)$, which are related to $H(t;f)$ by the ``pair ef{\kern0pt}ficiency'' $\epsilon_{IJ}(t;\hat\Omega)$:
\begin{eqnarray}
  \langle \hat{C}_{IJ}(t;f) \rangle & 
  \equiv & \epsilon_{IJ}(t;\hat\Omega,\vec\alpha) H(t;f)
  e^{-2\pi i f \tau_{IJ}} \\
  \langle\hat{P}_I(t;f)\rangle &
  \equiv & \epsilon_{II}(t;\hat\Omega,\vec\alpha) H(t;f) + N_I(t;f)
  \label{eq:autopower_def_body}
\end{eqnarray}
Here $\tau_{IJ}$ is the direction-dependent time delay between detector $I$ and detector $J$ and $N_I(t;f)$ is the noise power in detector $I$.
For additional details, including an expression for $\epsilon$, see \ref{app}.
It is also useful to define $\hat{P}_I'(t;f)$, the power in the $2n$ segments neighboring $t$:
\begin{equation}\label{eq:pprime_body}
  \hat{P}_I'(t_0;f) \equiv 
  \frac{1}{2n}\left[\sum_{t=t_0-n\delta t}^{t=t_0+n\delta t} \hat{P}_I(t;f) \right]- 
  \frac{1}{2n}  \hat{P}_I(t_0;f) .
\end{equation}
In this analysis we use $n=4$ neighboring segments on each side.

\section{An auto-power dif{\kern0pt}ference statistic}\label{Xi}
\begin{figure}[hbtp!]
  \centering
  \subfloat[]{\includegraphics[height=2.2in]{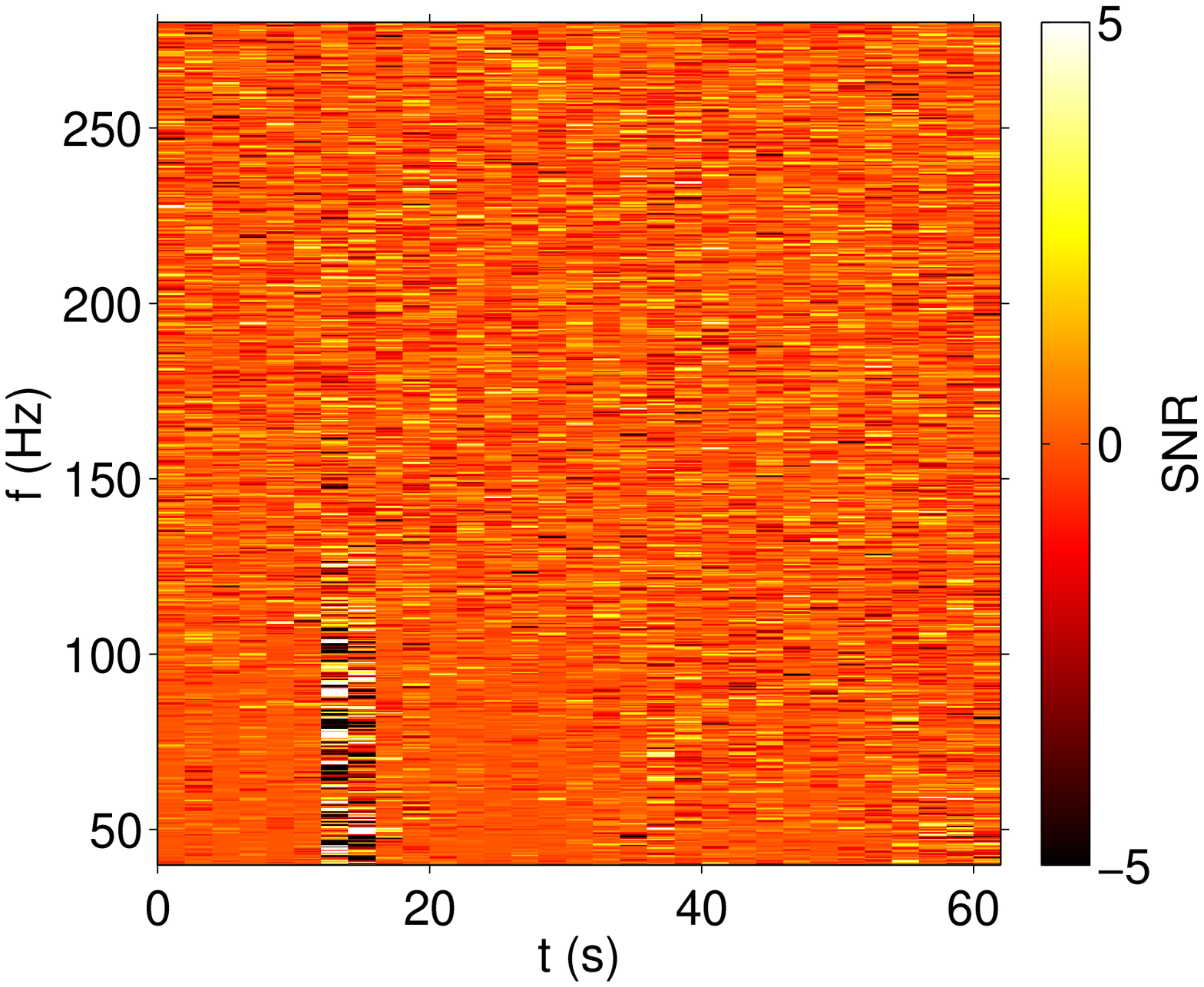}} \qquad
  \subfloat[]{\includegraphics[height=2.2in]{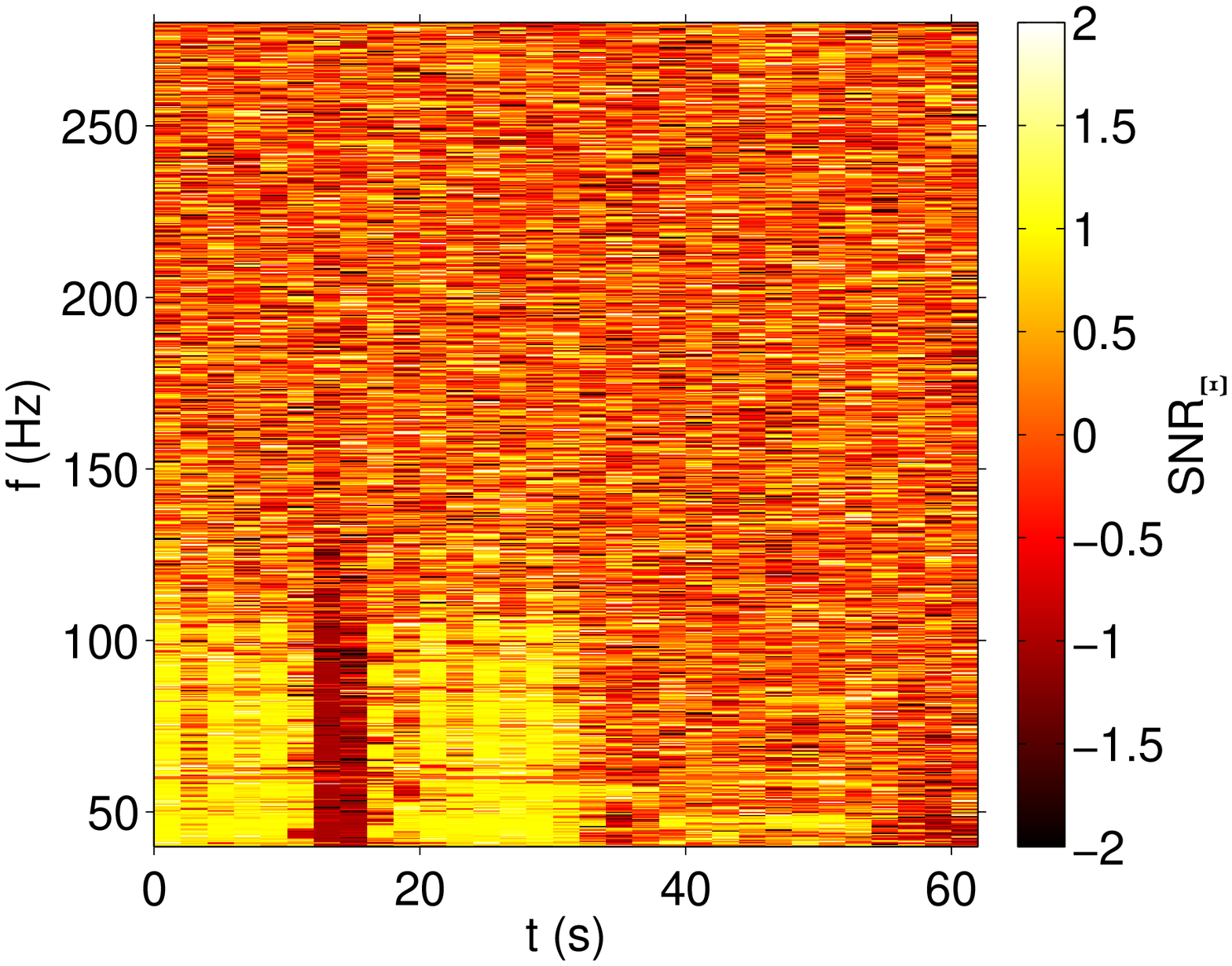}} \\
  \subfloat[]{\includegraphics[height=2.2in]{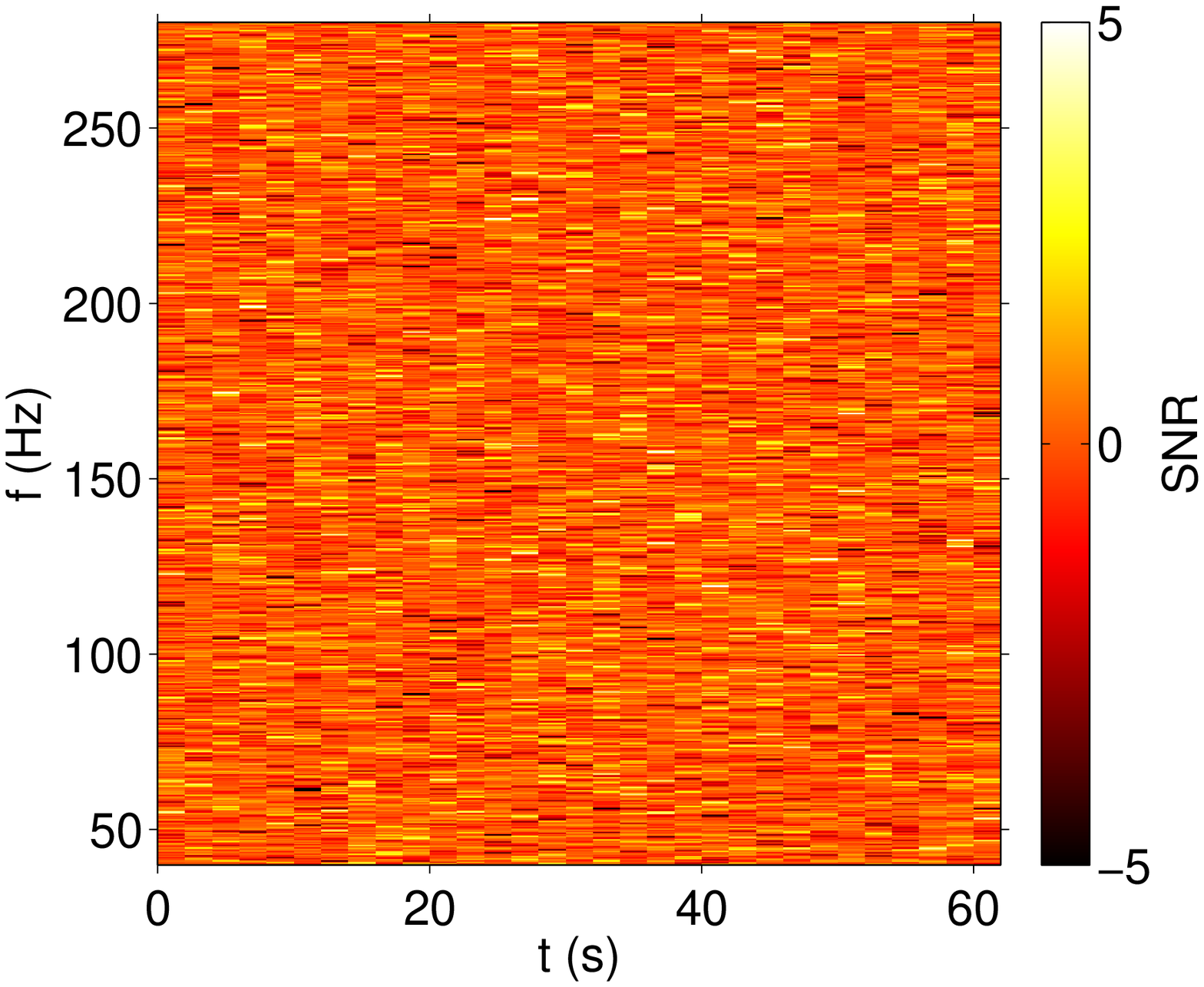}} \qquad
  \subfloat[]{\includegraphics[height=2.2in]{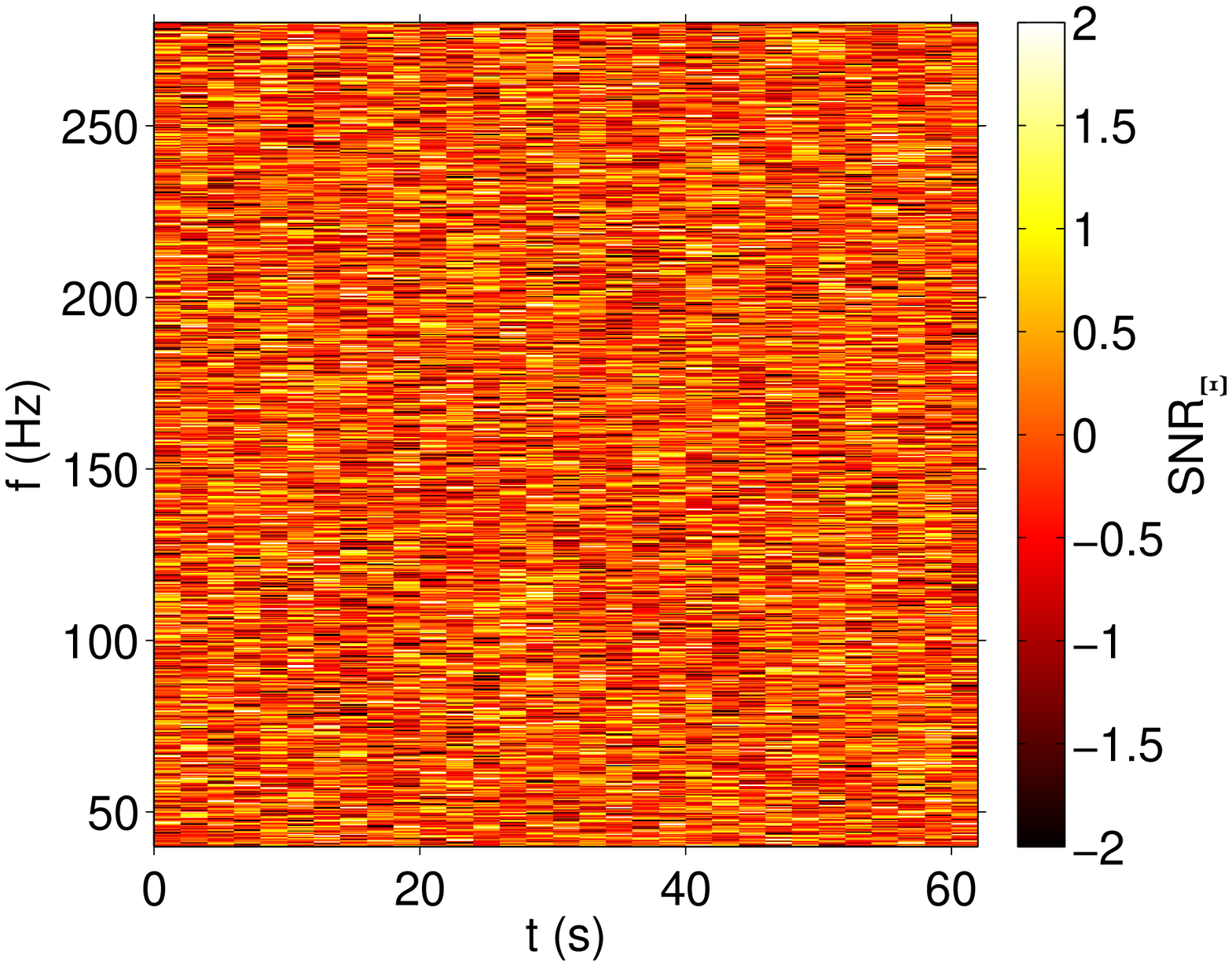}} \\
  \subfloat[]{\includegraphics[height=2.2in]{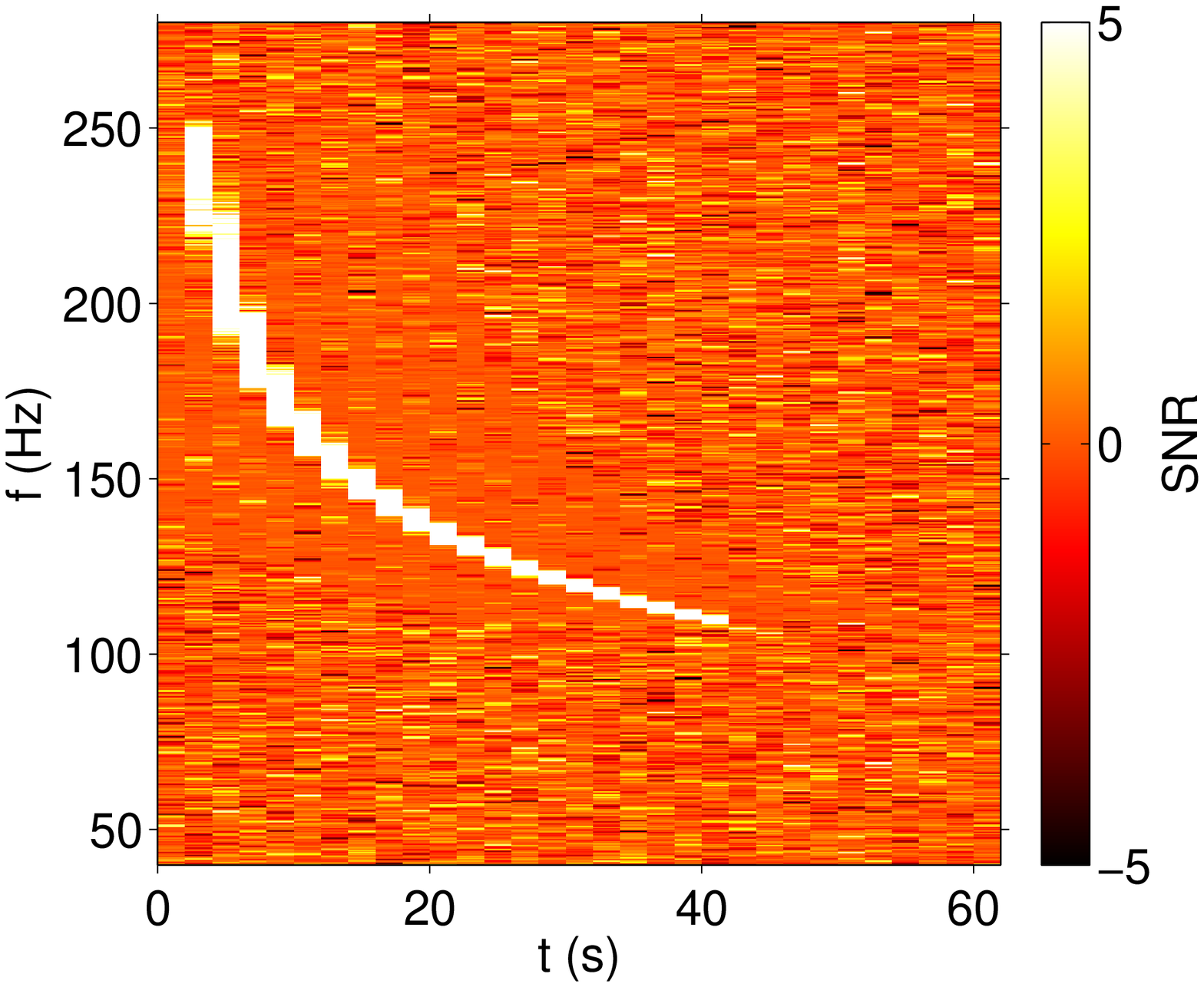}} \qquad
  \subfloat[]{\includegraphics[height=2.2in]{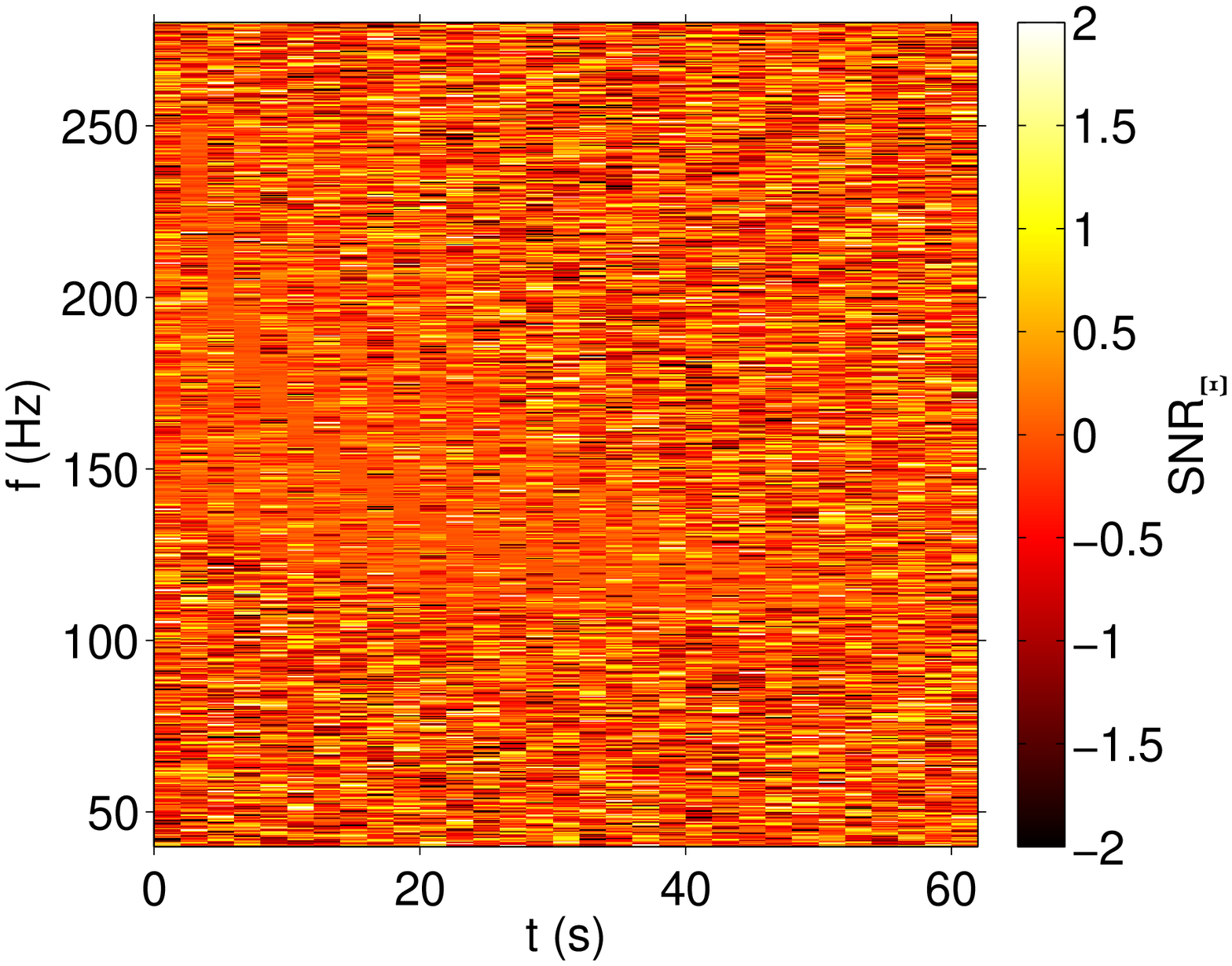}} \\
  \caption{$Ft$-maps of time-shifted LIGO S5 data. 
    The left-hand column shows cross-power $\text{SNR}$ while the right-hand column shows $\text{SNR}_\Xi$ for the same data.
    Top row: a likely glitch.
    Middle row: nearly stationary noise.
    Bottom row: stationary noise plus a simulated circularly-polarized accretion disk instability waveform ($d=\unit[5]{Mpc}$) (see~\cite{vanPutten,vanputten:01,vanputten:08}).}
  \label{fig:Xi_glitch}
\end{figure}

Since the noise and the signal are uncorrelated, the expectation value of $\hat{P}_I(t;f)$ is given by Eq.~\ref{eq:autopower_def_body}.
If we assume that $N_I(t;f)$ can be estimated by looking at neighboring segments of noise, (i.e., the noise is stationary), then we can construct an estimator for the observed auto-power in detector $I$ due to GWs:
\begin{equation}
  \frac{\hat{P}_I(t;f) - \hat{P}_I'(t;f)}{\epsilon_{II}} .
\end{equation}
We assume that there is no (or comparatively little) signal present in the same frequency bin during these neighboring times\footnote{This approximation works best for narrowband signals whose frequency varies significantly with time, as is the case for the examples shown here (see, e.g., Fig.~\ref{fig:Xi_glitch}). When the approximation is poor, e.g., for a monochromatic signal, then $\langle \hat{P}_I'(t;f)\rangle$ may include a significant GW component as well, though, $\hat\Xi(t;f)$ (defined in Eq.~\ref{eq:Xi_def}) will behave much the same way as it is still the case that $\langle\hat\Xi(t;f)\rangle=0$ by construction.} so that
\begin{equation}
  \langle \hat{P}_I'(t;f) \rangle \approx N_I(t;f) .
\end{equation}
Similarly, the GW auto-power in detector $J$ is:
\begin{equation}
  \frac{\hat{P}_J(t;f) - \hat{P}_J'(t;f)}{\epsilon_{JJ}} .
\end{equation}

We now construct a quantity, which represents the GW auto-power dif{\kern0pt}ference between detectors $I$ and $J$:
\begin{equation} \label{eq:Xi_def}
  \hat\Xi(t;f) \equiv 
  \frac{\hat{P}_I(t;f) - \hat{P}_I'(t;f)}{\epsilon_{II}} -
  \frac{\hat{P}_J(t;f) - \hat{P}_J'(t;f)}{\epsilon_{JJ}} .
\end{equation}
By construction, we expect that $\langle\hat\Xi(t;f)\rangle=0$ for well-behaved noise plus a signal that is well-modeled by the pair ef{\kern0pt}ficiencies $\epsilon_{II},\epsilon_{JJ}$.
We note that $|\hat\Xi(t;f)|$ is invariant under $I \leftrightarrow J$.

It is desirable to normalize $\hat\Xi(t;f)$ such that the new quantity is unitless with a near-unity variance.
The variance of $\hat\Xi(t;f)$ is given by:

\begin{equation}\label{eq:sigma_def}
  \fl \sigma^2_\Xi(t;f) 
  = \frac{ \langle \hat{P}_I(t;f) \rangle^2
    + \langle\hat{P}_I^{\prime} (t;f)\rangle^2}{\epsilon_{II}^2} + 
  \frac{\langle\hat P_J(t;f)\rangle^2
    + \langle\hat{P}_J^{\prime} (t;f)\rangle^2}{\epsilon_{JJ}^2}
  - \frac{2\,\epsilon_{IJ}^2}{\epsilon_{II}\,\epsilon_{JJ}}
  \left| \langle\hat Y(t;f) \rangle\right|^2 . 
\end{equation}
This motivates a normalization factor denoted $\hat\sigma_\Xi^2(t;f)$, which we choose to be
\begin{equation}
  \fl \hat\sigma^2_\Xi(t;f) 
  \equiv \frac{ \hat{P}_I^2(t;f)
    + \hat{P}_I^{\prime 2} (t;f)}{\epsilon_{II}^2} + 
  \frac{\hat P_J^2(t;f)
    + \hat{P}_J^{\prime 2} (t;f)}{\epsilon_{JJ}^2}
  - \frac{2\,\epsilon_{IJ}^2}{\epsilon_{II}\,\epsilon_{JJ}}
  \left| \hat Y(t;f) \right|^2 .
\end{equation}
We shall see below that this normalization provides an ef{\kern0pt}fective means of creating a unitless signal-to-noise ratio $\text{SNR}_\Xi\equiv\hat\Xi/\hat\sigma_\Xi$, which we can use to determine if the auto-power in two interferometers is consistent with a GW signal plus well-behaved (stationary) noise.
The $(t;f)$ dependence of $\text{SNR}_\Xi(t;f)$ is implicit.
Note that $\text{SNR}_\Xi$ is not equivalent to the cross-correlation statistic $\text{SNR}\equiv\hat{Y}/\sigma_Y$.

By considering Eqs.~\ref{eq:Xi_def} and~\ref{eq:sigma_def}, it is apparent that the qualitative behavior of $\text{SNR}_\Xi$ is dif{\kern0pt}ferent for signals $\langle\hat{Y}(t;f)\rangle>0$ and glitches $P_{I}(t;f)\gg P_{J}(t;f)$.
Loud glitches in detectors $I,J$ cause $\text{SNR}_\Xi\approx\pm1$ surrounded by $\text{SNR}_\Xi\approx\mp1$ (see, e.g., Fig.~\ref{fig:Xi_glitch}, top row).
Neighboring segments are af{\kern0pt}fected due to our noise estimation technique, which averages adjacent segments in time (see \ref{app}).
Loud GW signals, on the other hand, cause $\text{SNR}_\Xi\approx0$ surrounded by $\text{SNR}_\Xi\approx0$ with larger fluctuations in the neighboring segments.
This qualitative description of the $\text{SNR}_\Xi$ in the presence of a GW signal is demonstrated in Fig.~\ref{fig:Xi_dist_glitches} as well as the bottom row of $ft$-maps in Fig.~\ref{fig:Xi_glitch}.

\section{Glitch identification}\label{algorithm}
Having introduced the auto-power dif{\kern0pt}ference statistic $\text{SNR}_\Xi$, we now present an $\text{SNR}_\Xi$-based algorithm to identify glitches.
We use data collected from the LIGO S5 science run.
Our network consists of the two $\unit[4]{km}$ LIGO interferometers mentioned in Section~\ref{intro}.
The data are time-shifted by a duration greater than the GW travel time between H1 and L1 in order to wash out the presence of astrophysical signals.
To begin, we utilize $ft$-maps with $\unit[4]{s}\times\unit[0.25]{Hz}$ pixels.

\begin{figure}[hbtp!]
  \centering
  \subfloat[]{\includegraphics[height=2in]{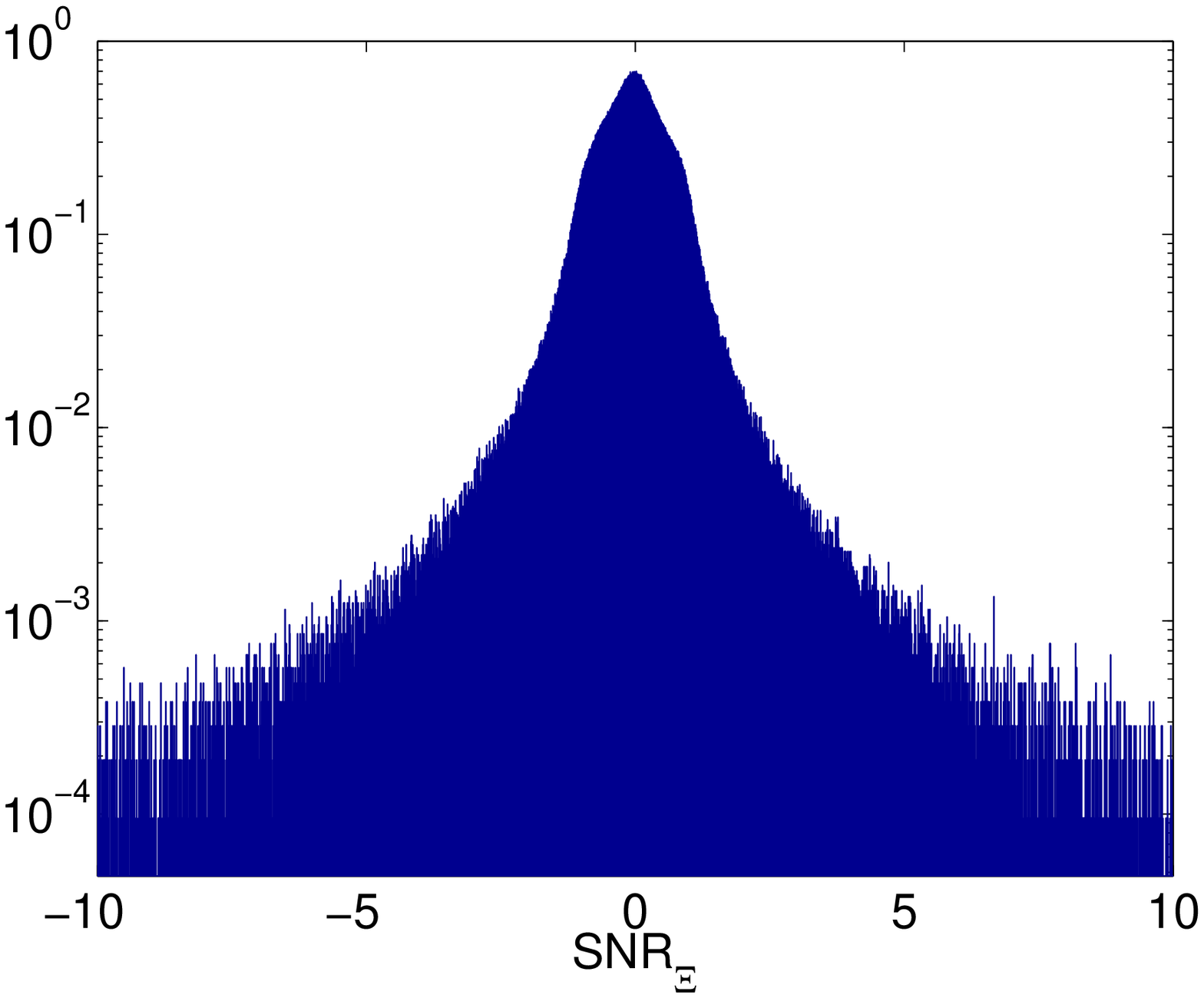}} \qquad
  \subfloat[]{\includegraphics[height=2in]{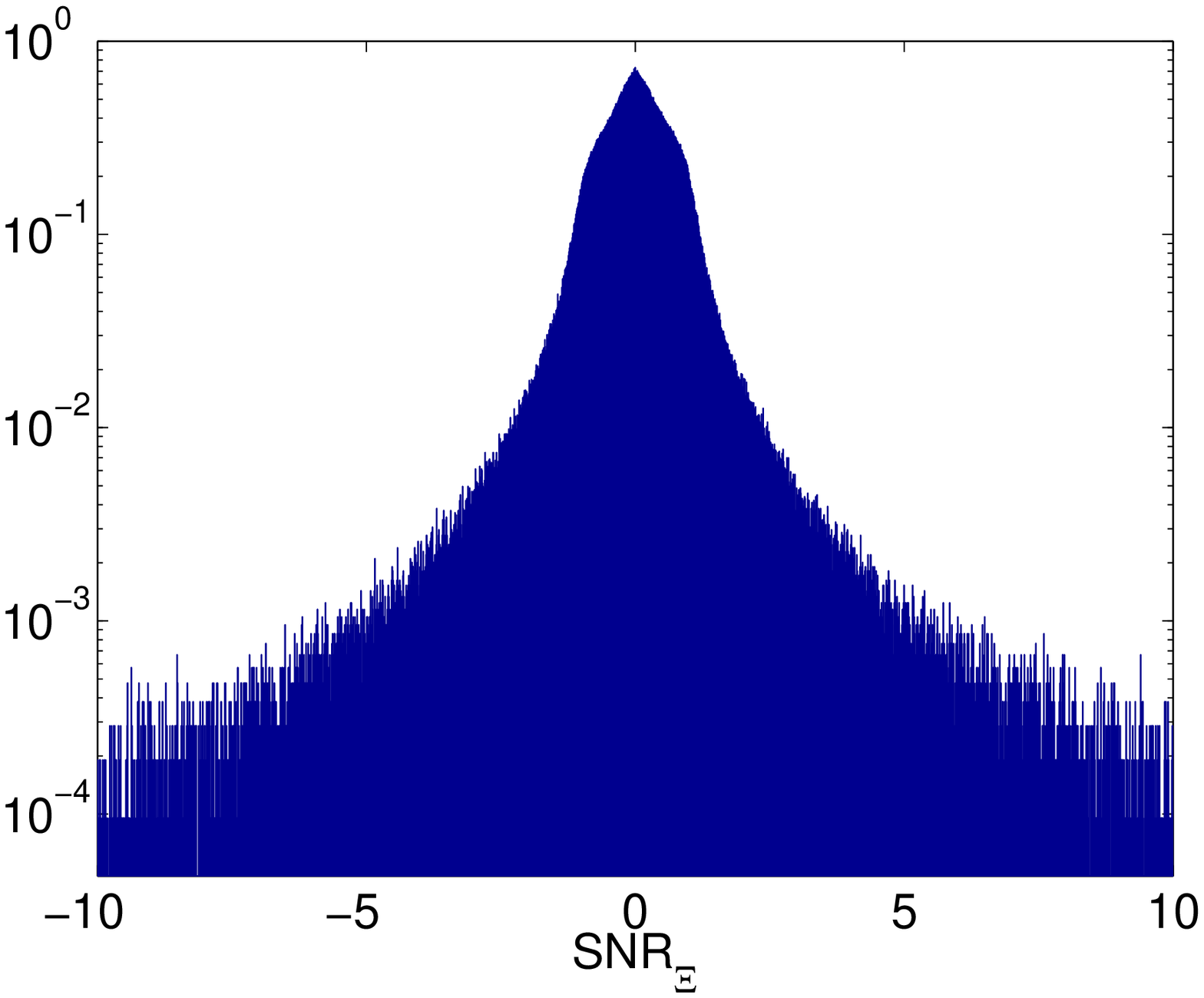}} \\
  \subfloat[]{\includegraphics[height=2in]{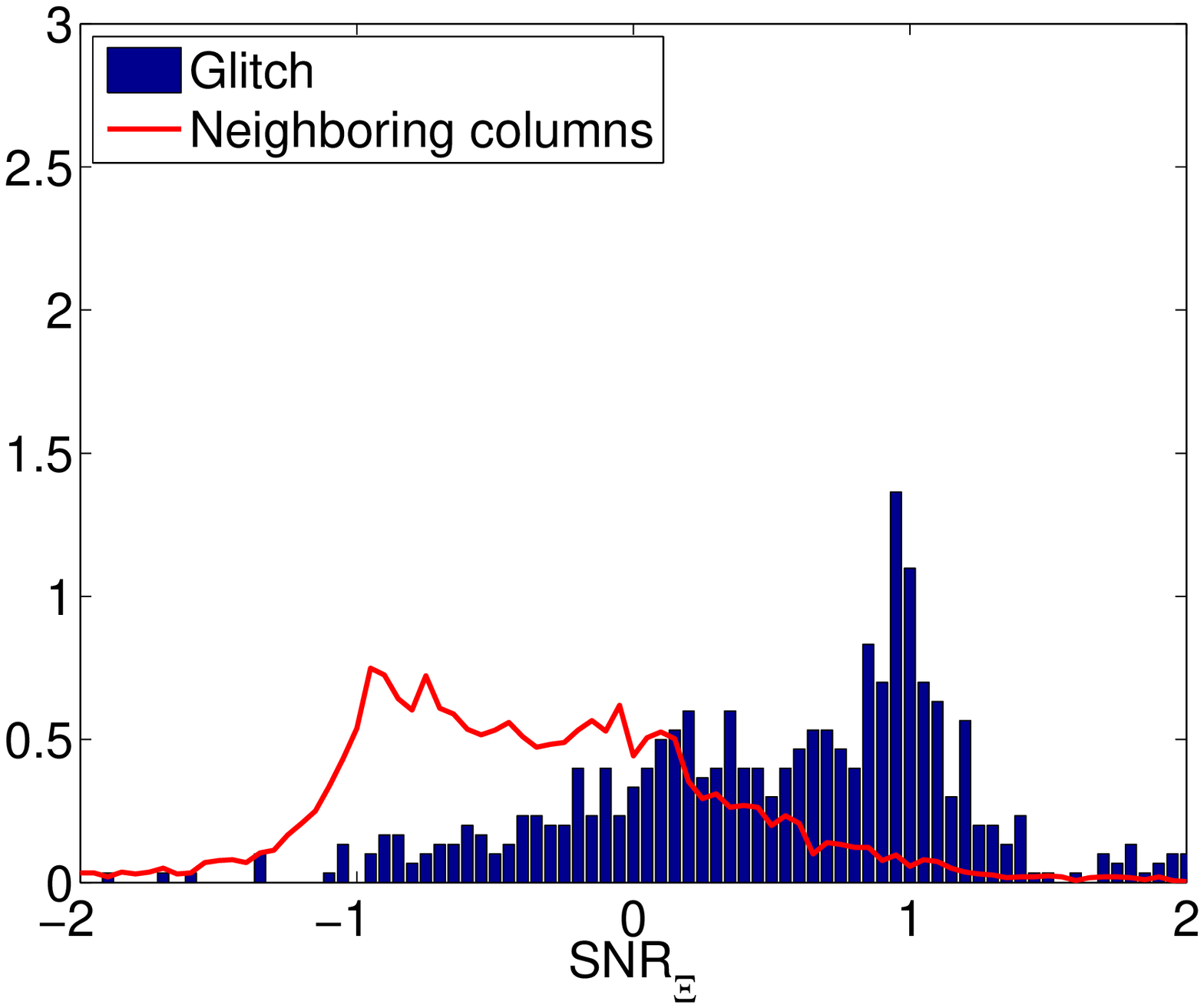}} \qquad
  \subfloat[]{\includegraphics[height=2in]{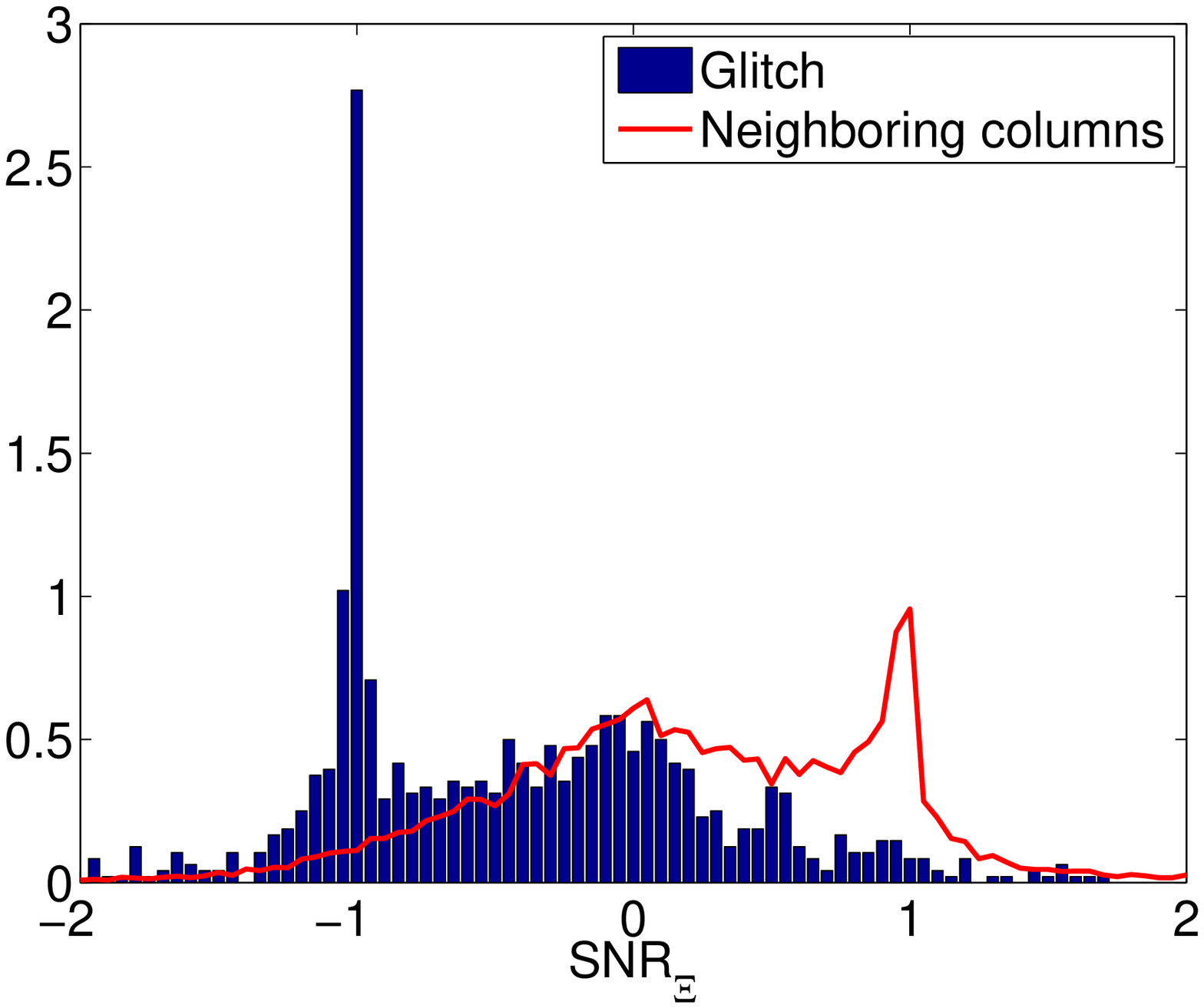}} \\
  \subfloat[]{\includegraphics[height=2in]{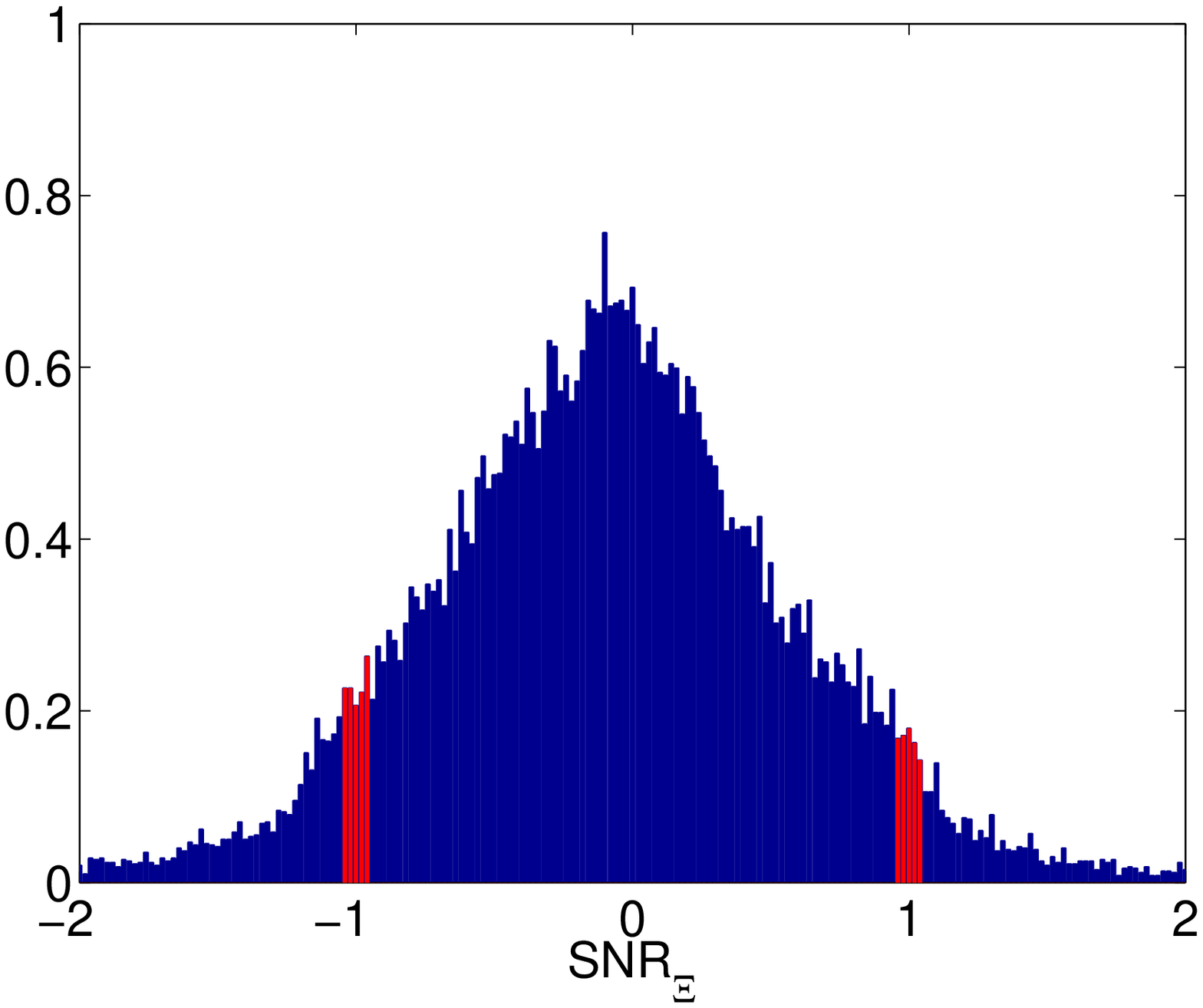}} \qquad
  \subfloat[]{\includegraphics[height=2in]{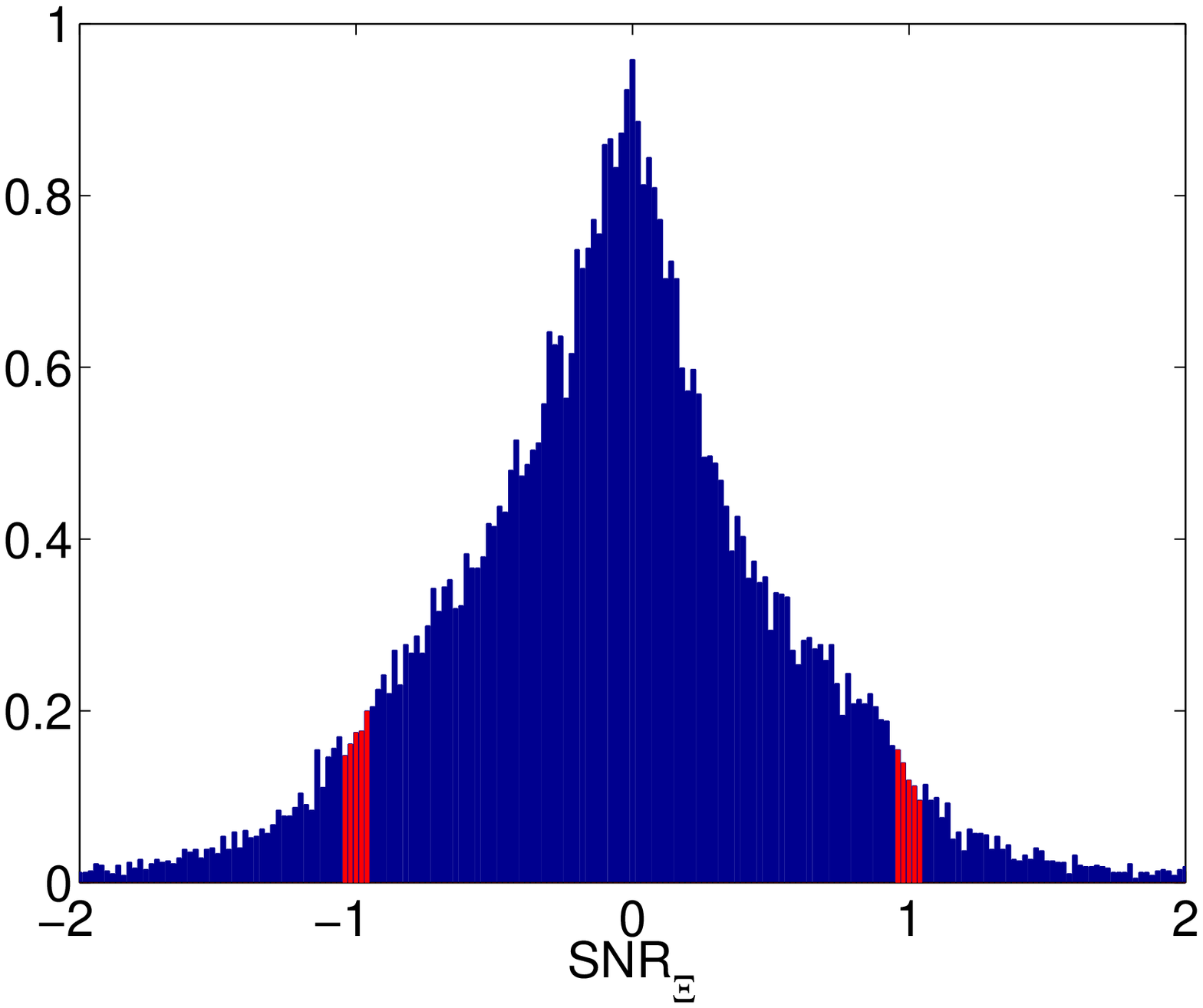}} \\
  \caption{
    Histograms of $\text{SNR}_{\Xi}$ $ft$-map pixels. 
    Top: $\unit[2200]{s}$ of Gaussian Monte Carlo noise (left) and well-behaved (nearly-Gaussian) time-shifted LIGO S5 data (right).
    Middle row: two examples of glitches in H1 (left) and L1 (right) each consisting of $\unit[2]{s}$ of data.
    These data segments were chosen to illustrate examples of strong glitches.
    Bottom: $\unit[40]{s}$-long circularly-polarized accretion disk instability injections recovered with an unpolarized filter at $\unit[30]{Mpc}$ (left) and at $\unit[5]{Mpc}$ (right).
    The red bars indicate $0.95<|\text{SNR}_\Xi|<1.05$.  
    The $x$-axis range dif{\kern0pt}fers between rows due to the dif{\kern0pt}ferent amount of data being analyzed in each case.
  }
  \label{fig:Xi_dist_glitches}
\end{figure}

In Fig.~\ref{fig:Xi_dist_glitches} we show the distribution of $\text{SNR}_{\Xi}$ for well-behaved noise (top), glitchy noise (middle) and a simulated accretion disk instability (ADI) GW signal (see Section~\ref{toy}) injected on top of Gaussian noise (bottom).
``Well-behaved'' means that there are no obvious high-level glitches visible in an $ft$-map of $\text{SNR}$, which is to say that the data approximate stationary Gaussian noise.
As examples of glitchy noise, we utilize data from two extreme glitches; one from H1 and one from L1.
As stated in Section~\ref{Xi}, we observe that glitches cause an excess of pixels near $|\text{SNR}_\Xi|=1$.
However, if we simply flag segments with  $|\text{SNR}_\Xi|\approx1$, we will throw out more data than necessary because segments neighboring a glitch also exhibit $|\text{SNR}_\Xi|\approx1$.

To discriminate between the glitch segment and its neighbors, we define an additional metric, the {\em auto-power stationarity ratio}:
\begin{equation}
  R_I(t) = \frac{1}{N_f}\sum_f \frac{\hat{P}_{I}(t;f)}{\hat{P}_{I}'(t;f)} .
\end{equation}
Here $N_f$ is the number of frequency bins.
We expect segments with a glitch to have $R_I(t)\gtrsim1$ whereas neighboring segments should have $R_I(t)<1$.
(Of course, GW signals can also lead to $R_I(t)\gtrsim1$ so it is necessary to use $R$ in conjunction with $\text{SNR}_\Xi$ in order to separate glitches from GW events.)
Glitches are unlikely to occur in two interferometers at the same time.

Now we are ready to devise our glitch likely flag.
A data segment (or equivalently, an $ft$-map column) is identified as {\em glitch-like} if either of the following criteria are satisfied:
\numparts
\begin{eqnarray}\label{eq:criteria}
  && \fl \text{$\geq 2.7\%$ of pixels have $0.95<\text{SNR}_\Xi<1.05$ and $R_I(t)>2$ and $R_J(t)\leq2$.}\label{eq:flag_params1} \\
  && \fl \text{$\geq 2.7\%$ of pixels have $-1.05<\text{SNR}_\Xi<-0.95$ and $R_J(t)>2$ and $R_I(t)\leq2$.}\label{eq:flag_params2}
\end{eqnarray}
\endnumparts
These parameters are chosen primarily to optimize the ef{\kern0pt}ficiency of our algorithm at rejecting glitches, though some fine tuning is necessary to ensure the safety of a particular signal model.
In this case, the parameters are adjusted for the ADI model (see Fig.~\ref{fig:scatter}), but we shall see that they are also ef{\kern0pt}fective for a very dif{\kern0pt}ferent signal model (based on accretion disk fragmentation) in Section~\ref{toy}.
 Before we continue, it will be useful to define ${\cal F}$ as the ratio of the number of pixels at some time $t$ satisfying the criterion $0.95<\left|\text{SNR}_\Xi\right|<1.05$ to the total number of pixels at time $t$.
Note that ${\cal F}$, by definition, must take on discrete values.

In order for this to be an ef{\kern0pt}fective flag, it must not only identify glitch-like structures in the data, but it should also have a low \emph{false glitch rate}.
We define the false glitch rate as the fraction of Gaussian-noise segments (containing no glitches) flagged as glitch-like per unit time.
Using simulated Gaussian data, we estimate a false glitch rate of $\lesssim\unit[1\times10^{-3}]{day^{-1}}$.
This false glitch rate is calculated for a frequency range between $\unit[100-250]{Hz}$ consisting of 150 pixels, a range suitable for the ADI model that we will use to test this algorithm (see Section~\ref{toy}).

\begin{figure}[hbtp!]
  \centering
  \includegraphics[height=2.5in]{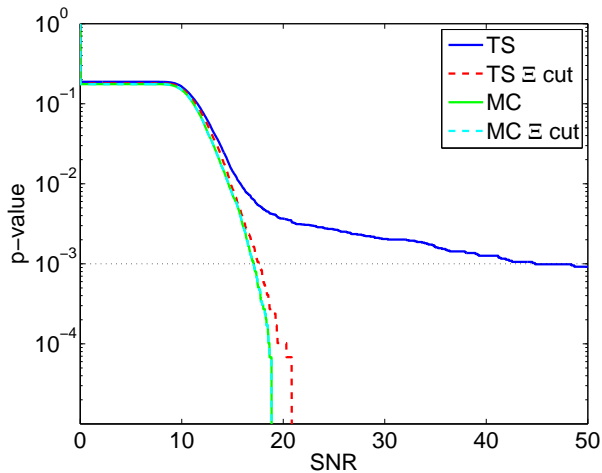} 
  \caption{
    Plot of $p$-value vs.\ $\text{SNR}$ for a density-based search algorithm~\cite{burstCluster} applied to time-shifted LIGO S5 data (TS) and Gaussian Monte Carlo data (MC), with and without our $\text{SNR}_{\Xi}$-based glitch cut applied, for $\unit[4]{s}\times\unit[0.25]{Hz}$ pixels in a frequency band of $\unit[100-250]{Hz}$.
    The $\text{SNR}_{\Xi}$-based glitch cut improves the sensitivity at $p=0.1\%$ (marked with a black dotted line) by more than a factor of two.
The asymptotic $p$-value at low SNR is the probability that any above-threshold cluster is identified.
  }
  \label{fig:bc_bknd_study}
\end{figure}

To determine the ef{\kern0pt}fectiveness of our flag, we perform a background study comparing time-shifted data (containing stationary noise and glitches) with Monte Carlo (stationary noise) with and without flagged data removed.
An ef{\kern0pt}fective flag eliminates high-$\text{SNR}$ events from the tail of the distribution, thereby creating better agreement between time-shifted and Monte Carlo data.
We utilize a density-based search algorithm~\cite{burstCluster} to analyze $\unit[12]{s}\times\unit[150]{Hz}$ $ft$-maps with $\unit[4]{s}\times\unit[0.25]{Hz}$ pixels\footnote{The algorithm is a modified version of {\tt BurstCluster} by Peter Kalmus and Rubab Khan created for the LIGO Flare Pipeline (see~\cite{ligo_sgr,ligo_sgr:08}).}.
We focus on a frequency range of $\unit[100-250]{Hz}$ in order to study the ADI signal discussed in Section~\ref{toy} (see Fig.~\ref{fig:Xi_glitch}, bottom row).
In Fig.~\ref{fig:bc_bknd_study} we plot $p$-value (false alarm probability) vs.\ $\text{SNR}$ for Monte Carlo and time-shifted data with and without the glitch-likely flag.
The results indicate a significant improvement in the agreement between time-shifted and Monte Carlo data with the application of the flag.
The required $\text{SNR}$ for a $p=0.1\%$ event is reduced more than two-fold through the use of the glitch identification flag.

\begin{figure}[hbtp!]
  \centering
  \subfloat[]{\includegraphics[height=2.3in]{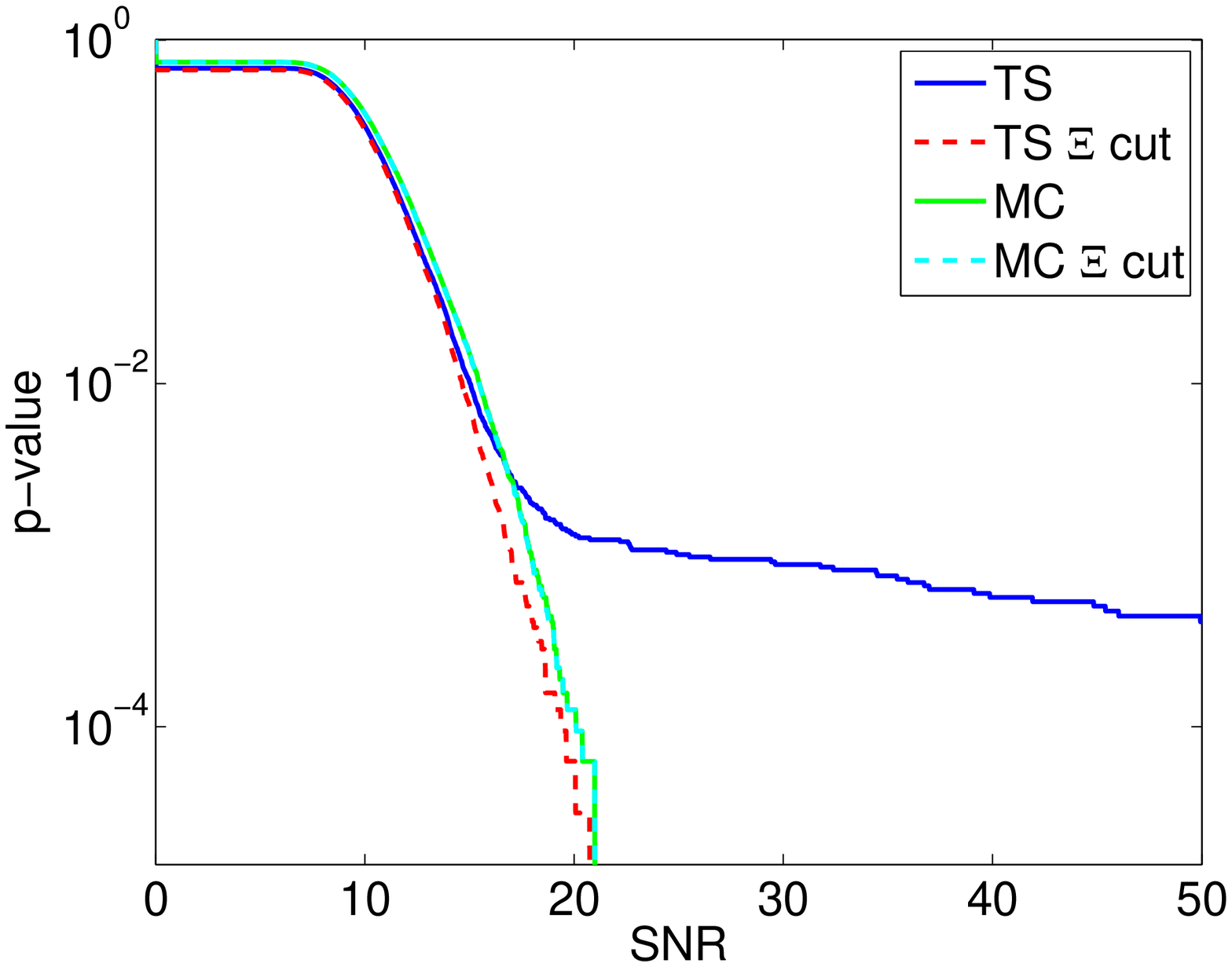}} \hspace{7pt}
  \subfloat[]{\includegraphics[height=2.3in]{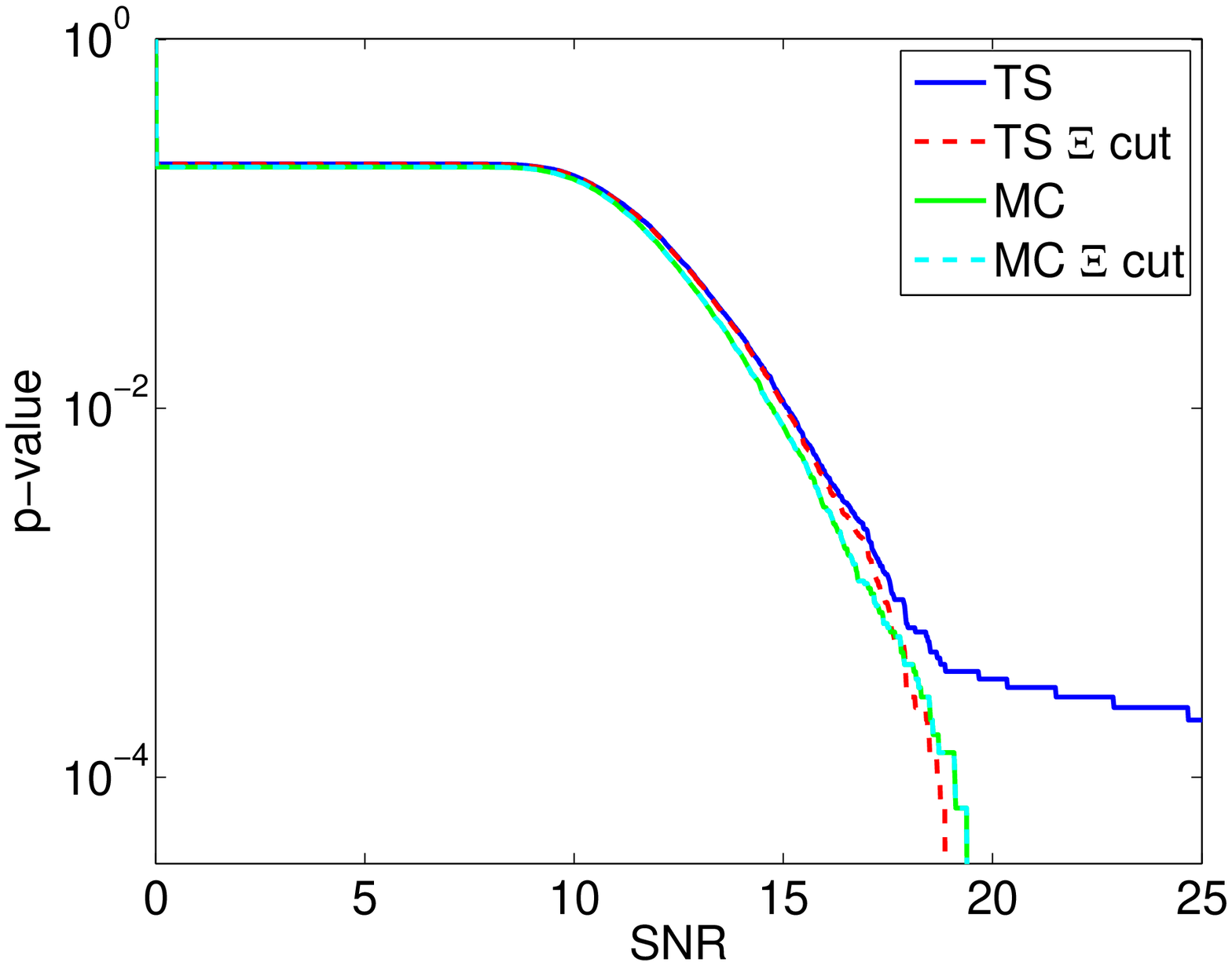}} \\
  \subfloat[]{\includegraphics[height=2.3in]{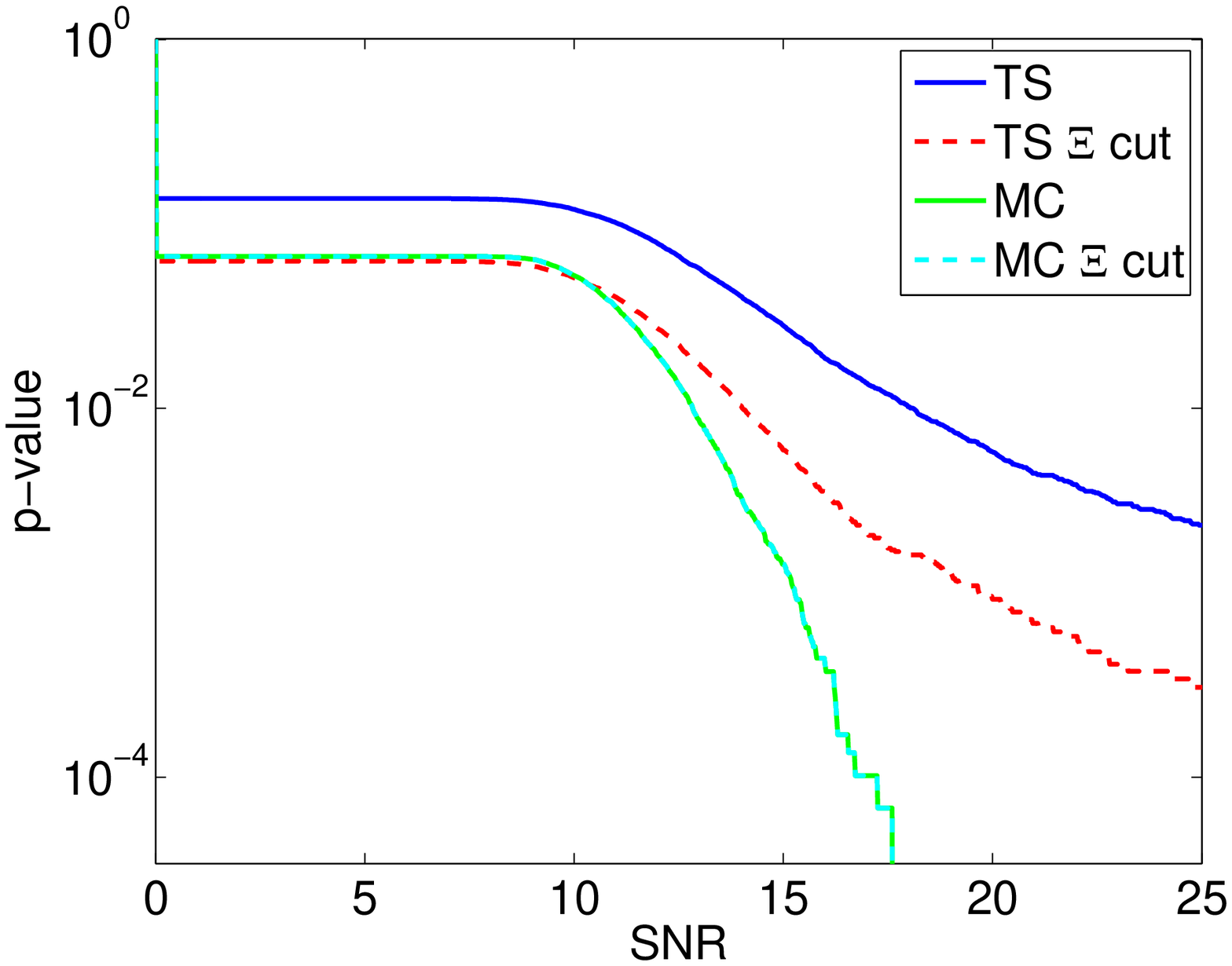}} \hspace{7pt}
  \subfloat[]{\includegraphics[height=2.3in]{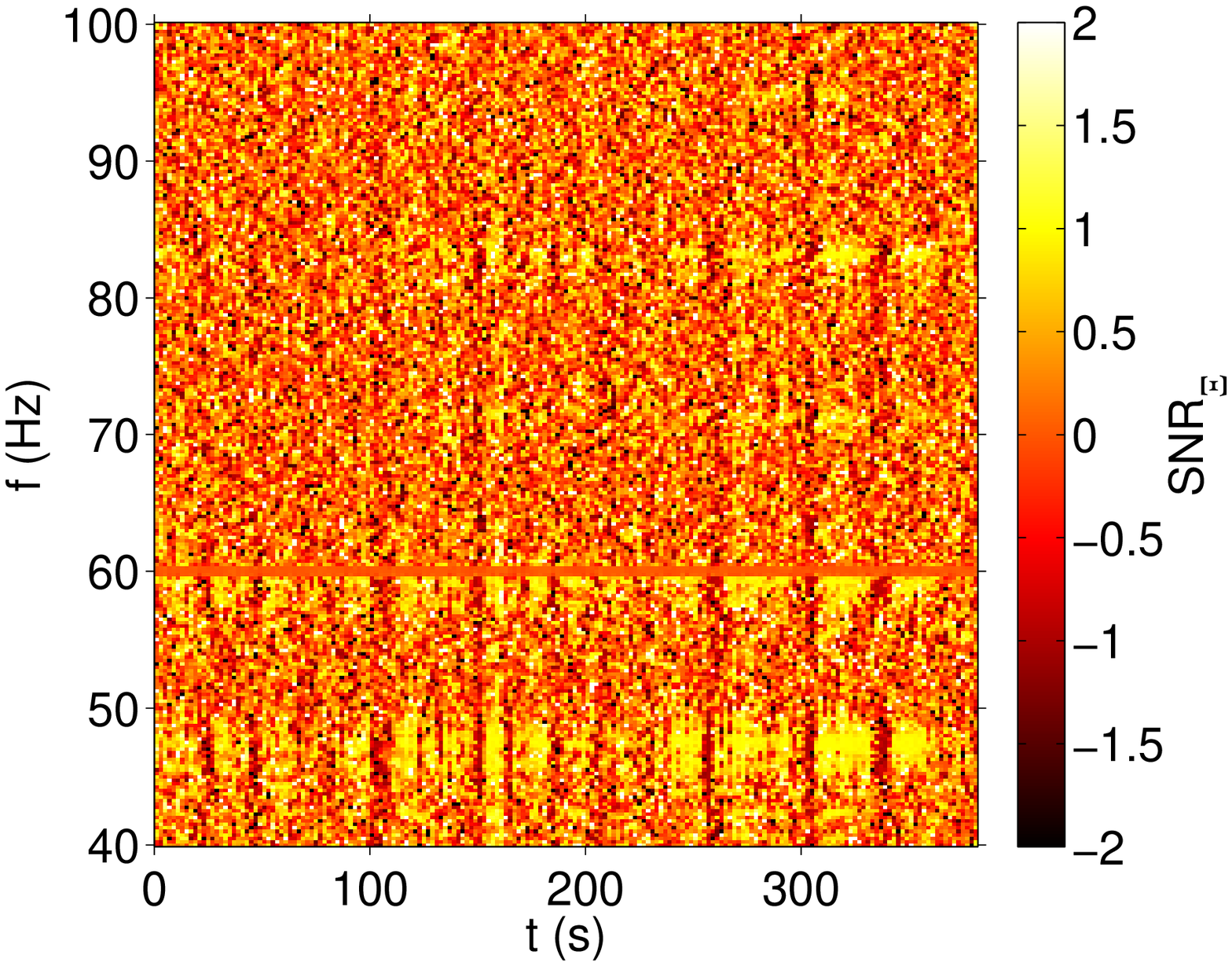}} \\
  \caption{
    Top left: plot of $p$-value vs.\ $\text{SNR}$ for the $\unit[100-250]{Hz}$ band using $\unit[1]{s}\times\unit[1]{Hz}$ pixels (the asymptotic $p$-value at low SNR dif{\kern0pt}fers from the $\unit[4]{s}\times\unit[0.25]{Hz}$ case since we tune the clustering algorithm dif{\kern0pt}ferently for dif{\kern0pt}ferent segment durations). The relatively good agreement between Monte Carlo (MC) and time-shifted data (TS) suggests that even short ${\cal O}(\unit[1]{s})$ segments of cross-correlated data can be ef{\kern0pt}fectively cleaned with our glitch identification flag.
    Top-right: plot of $p$-value vs.\ $\text{SNR}$ for the $\unit[375-525]{Hz}$ band using $\unit[4]{s}\times\unit[0.25]{Hz}$ pixels.  This higher frequency band exhibits good agreement between Monte Carlo and time-shifted data due to the nearly stationary noise associated with higher frequencies. 
    Bottom-left:  plot of $p$-value vs.\ $\text{SNR}$ for the $\unit[40-100]{Hz}$ band using $\unit[4]{s}\times\unit[0.25]{Hz}$ pixels. While the cut dramatically improves the agreement between Monte Carlo and time-shifted data, significant disagreement remains, possibly due to non-stationary noise associated with this band.
    Bottom-right: an $ft$-map of $\text{SNR}_\Xi$ for time-shifted LIGO S5 data demonstrating the non-stationary noise sometimes associated with low frequencies.  The $\unit[60]{Hz}$ line is masked.
}
  \label{fig:1s_bknd_study}
\end{figure}

Having demonstrated the ef{\kern0pt}ficacy of our glitch identification algorithm for the case of $\unit[4]{s}\times\unit[0.25]{Hz}$ pixels in the $\unit[100-250]{Hz}$ band chosen to study the ADI model (see Section~\ref{toy}), we now consider a few other cases.
An exhaustive exploration of the domain of utility for this algorithm is beyond our present scope.
Rather, we aim to highlight both the promise and the limitations of this technique by considering a few more special cases.
In the top-left panel of Fig.~\ref{fig:1s_bknd_study}, we plot $p$-value vs.\ $\text{SNR}$ for $\unit[1]{s}\times\unit[1]{Hz}$ pixels in the same $\unit[100-250]{Hz}$ frequency band used in Fig.~\ref{fig:bc_bknd_study}.
Based on the agreement between time-shifted and Monte Carlo data, we conclude that even relatively short segments of data can be ef{\kern0pt}fectively cleaned with the glitch identification algorithm so as to achieve good agreement between Monte Carlo and time-shifted data.

In the top-right plot, we show the case of $\unit[4]{s}\times\unit[0.25]{Hz}$ pixels in a higher frequency band: $\unit[375-525]{Hz}$.
Again we observe good agreement between time-shifted and Monte Carlo data, though, this is not surprising since this higher frequency band is typically dominated by nearly stationary noise.
On the bottom-left, we show the case of $\unit[4]{s}\times\unit[0.25]{Hz}$ pixels in a lower frequency band: $\unit[40-100]{Hz}$.
While the agreement between time-shifted and Monte Carlo data is improved with the glitch identification algorithm, significant disagreement remains due to non-stationary noise, which is more common at lower frequencies.
An $\text{SNR}_\Xi$ $ft$-map from a period of noisy low-frequency data is included in the bottom-right panel, which indicates that this effect may be due to quasi-continuous broadband noise rather than infrequent glitches. 
It is possible that the inclusion of additional vetoes utilizing physical environmental monitors such as microphones and seismometers may help achieve better agreement between time-shifted data in this band and Monte Carlo noise.

\begin{figure}[hbtp!]
  \centering
  \subfloat[]{\includegraphics[height=2.3in]{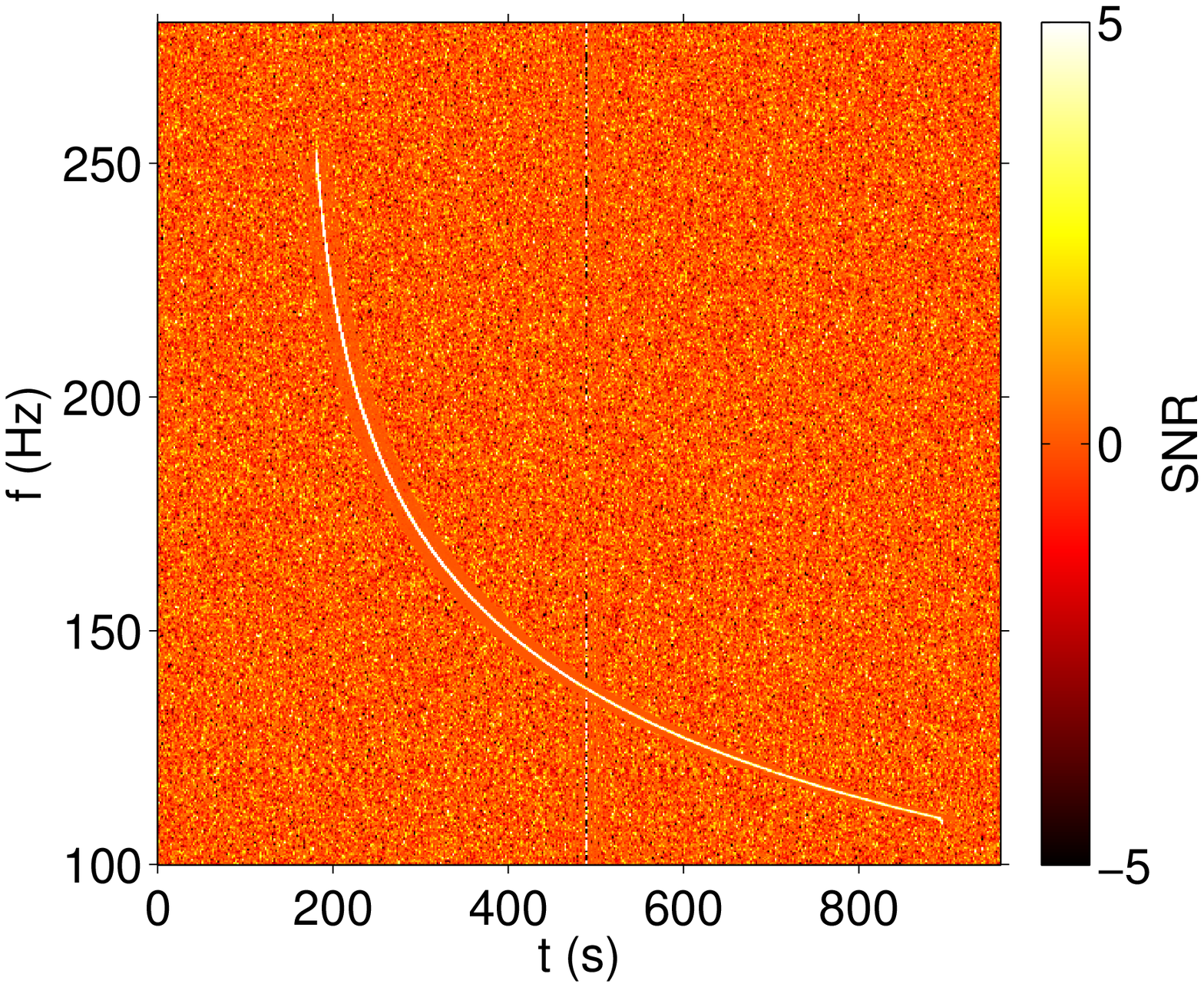}} \hspace{7pt}
  \subfloat[]{\includegraphics[height=2.3in]{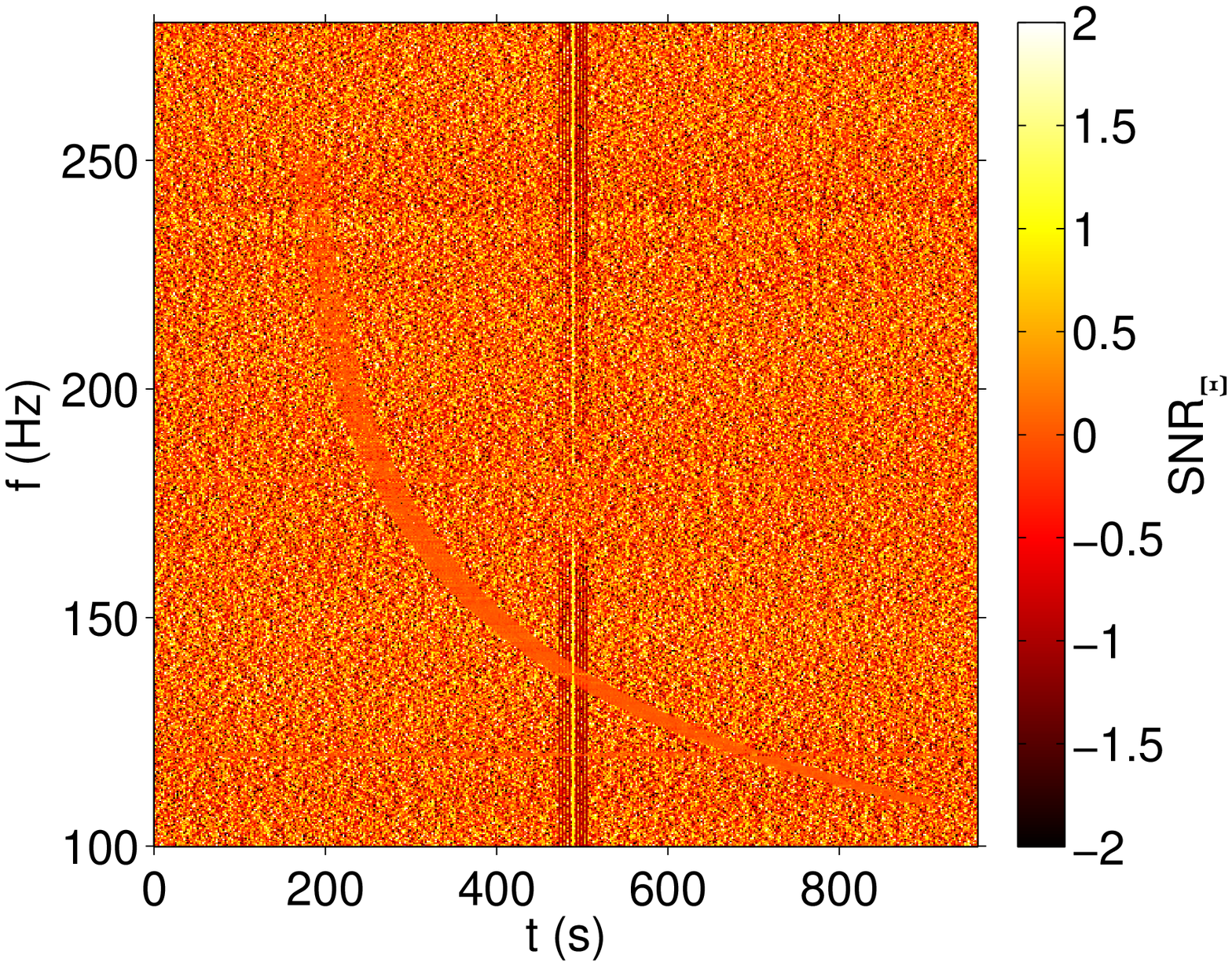}} \\
  \subfloat[]{\includegraphics[height=2.3in]{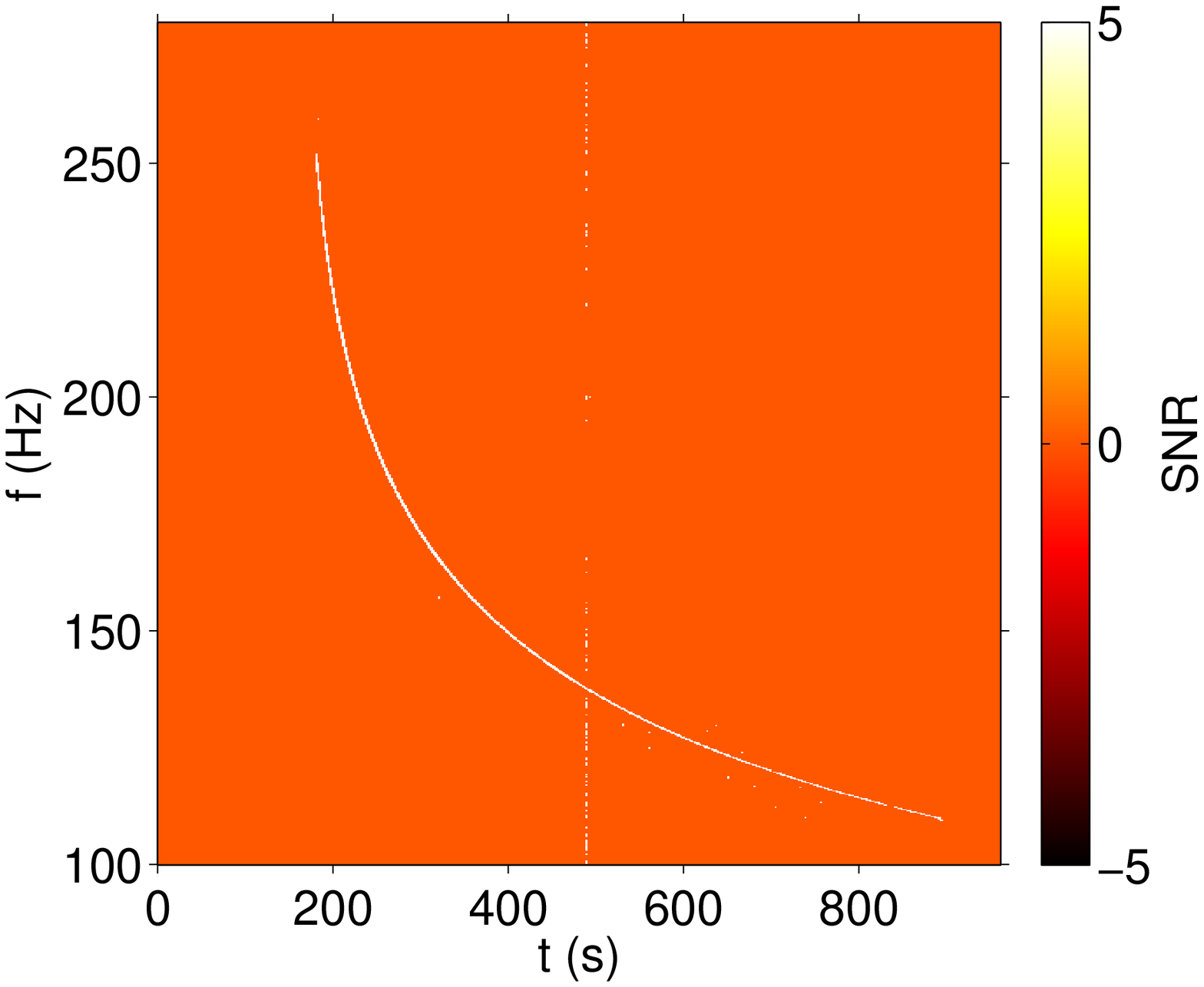}} \hspace{7pt}
  \subfloat[]{\includegraphics[height=2.3in]{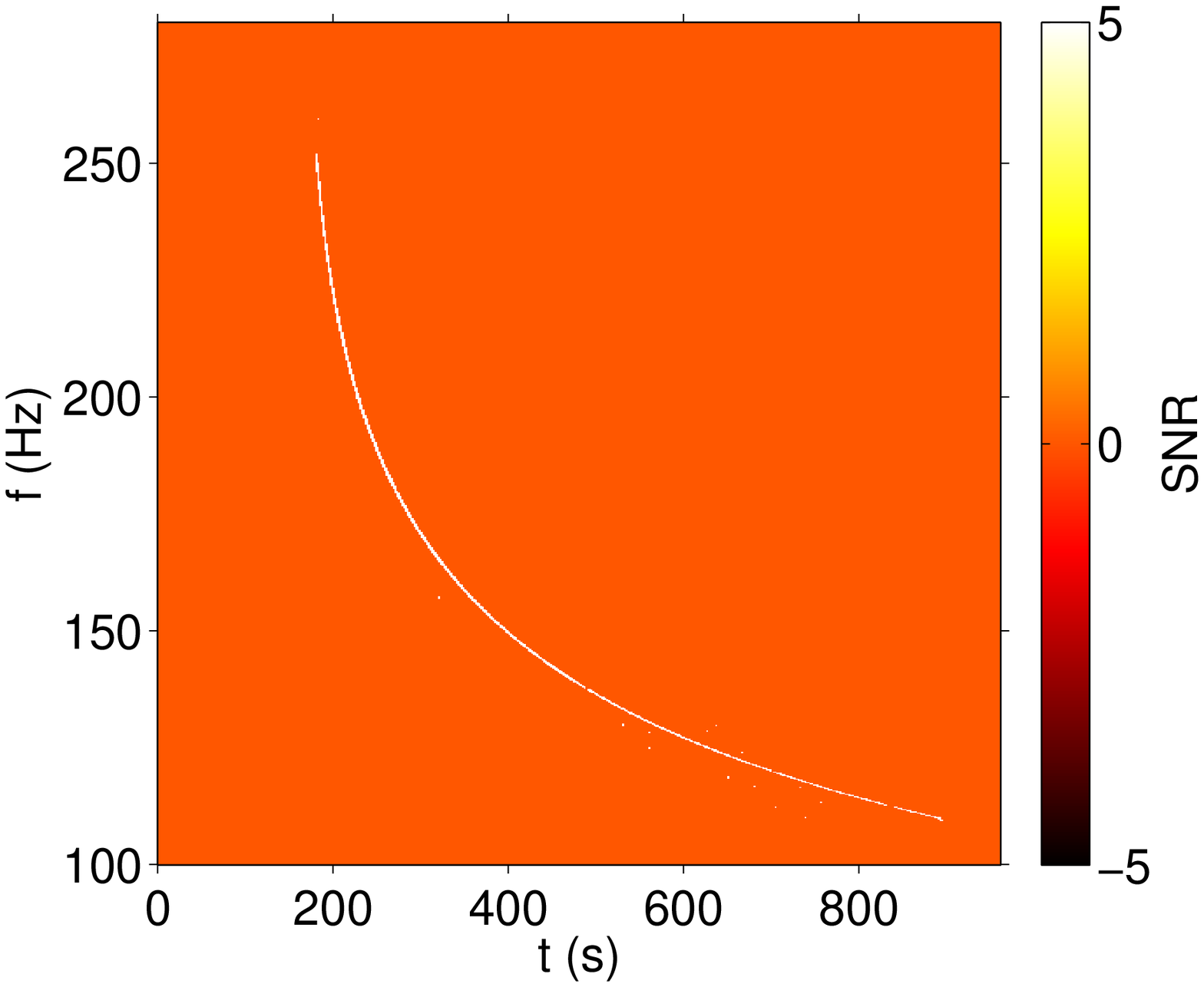}} \\
  \caption{$Ft$-maps of an ADI software injection in time-shifted data containing a glitch (near $t\approx\unit[490]{s}$).  Top-left is $\text{SNR}$ and top-right is $\text{SNR}_{\Xi}$. The bottom plots show the recovered track without (left) and with (right) the glitch identification algorithm.  The glitch identification algorithm excludes the glitch (visible as a vertical column of bright pixels) from the reconstructed track.}
  \label{fig:long_inj}
\end{figure}

Finally, in Fig.~\ref{fig:long_inj}, we demonstrate how the glitch identification algorithm can be used to improve the accuracy of a long reconstructed signal by removing one or more glitchy segments.
Motivated by models of long GW transients, which may last for hundreds of seconds or longer (e.g.,~\cite{corsi:09b,stella05}), we consider a 700 s-long ADI waveform.
We inject the waveform into time-shifted Gaussian noise during a period with a known glitch (visible as a vertical column around $t\approx\unit[490]{s}$).
Using the density-based clustering algorithm described above, we recover the track without (bottom-left) and with (bottom-right) the glitch identification algorithm applied.
The glitch identification algorithm correctly identifies the glitch, which is therefore excluded from the reconstructed event.
This demonstrates not only that the glitch identification algorithm improves the accuracy of a reconstructed track, but also that it is in theory possible to observe a GW event disrupted by a glitch.
While this possibility is discussed in~\cite{stamp_pem}, this is (to our knowledge) the first time that a method has been proposed for removing pieces of glitchy data from the middle of a GW trigger using only strain data.

\section{Toy model waveforms}\label{toy}
In order to demonstrate our glitch identification algorithm, we utilize a toy model~\cite{lucia} for a narrowband signal from an accretion disk instability, which can take place during the collapsar death of a star and may therefore be associated with long gamma-ray bursts (see also~\cite{vanPutten,vanputten:01}).
In this ``suspended accretion'' model, a spinning black hole (with mass $M$ and parameterized by a dimensionless spin parameter $a^\star$) is surrounded by a torus (mass $m$).
The spinning black hole drives magneto-hydrodynamical turbulence in the torus, which causes it to form clumps with mass given by $\epsilon \, m$.
These clumps emit elliptically polarized narrowband gravitational radiation for a duration of ${\cal O}(\unit[10-100]{s})$ as the central black hole transfers its angular momentum to the clumps.
This toy model provides a useful test of our algorithm because we expect many sources of long GW transients to be both narrowband and elliptically polarized \cite{stamp}.
We use $M=10M_\odot$, $m=1.5M_\odot$, $\epsilon=0.1$ and $a^\star=0.95$ to create the $\approx\unit[40]{s}$ waveform used here.
A spectrogram of this waveform can be seen in the bottom-left panel of Fig.~\ref{fig:Xi_glitch}.
For more details see~\cite{lucia}.

In addition to the ADI model, we also consider an accretion disk fragmentation model from~\cite{lucia}.
In this model, an accretion disk associated with a long gamma-ray burst forms clumps through helium photo-disintegration~\cite{piro}.
These clumps inspiral into the remnant black hole, creating a chirp-like GW signal.

\begin{figure}[hbtp!]
  \centering
  \subfloat[]{\includegraphics[height=2.3in]{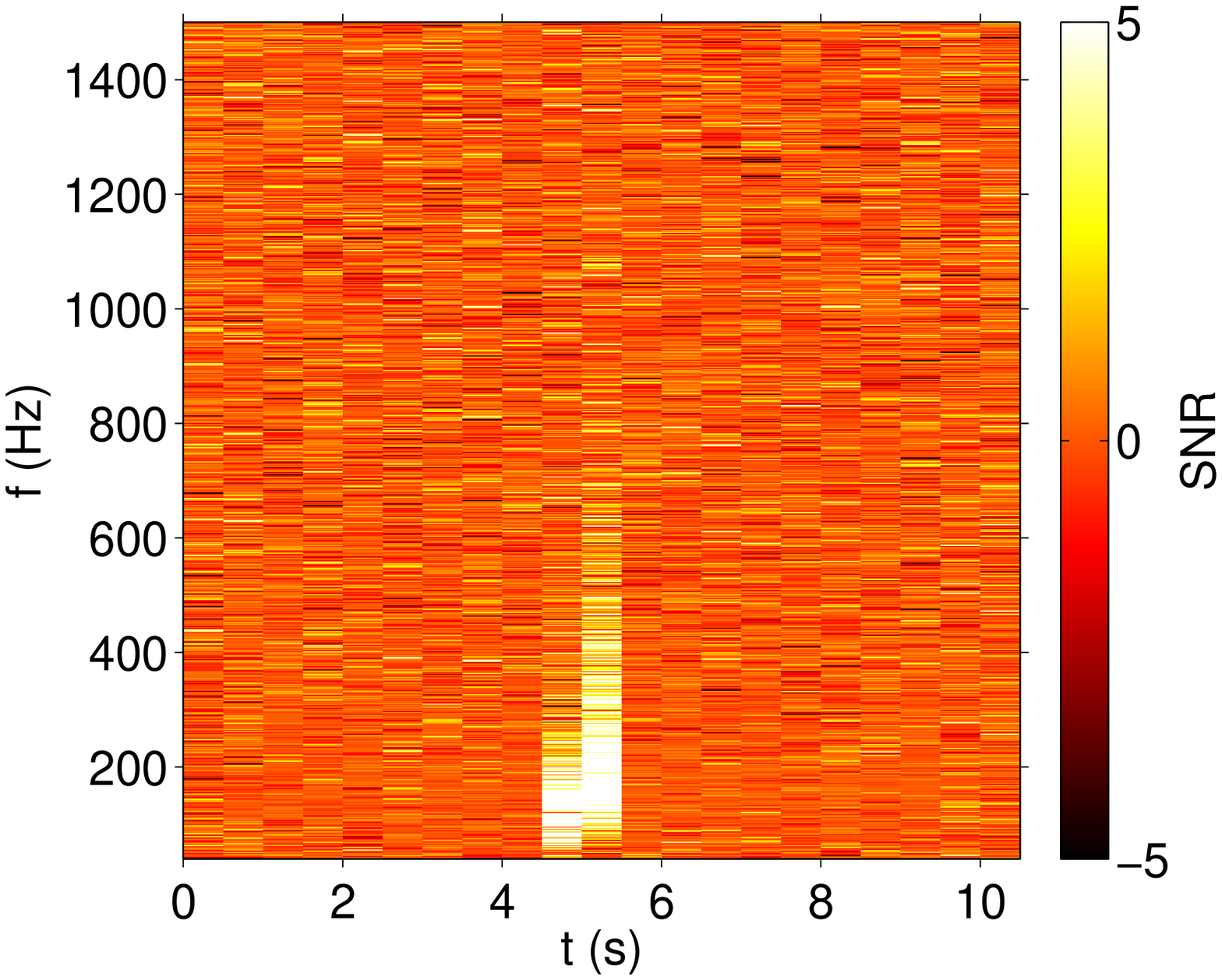}} \hspace{7pt}
  \subfloat[]{\includegraphics[height=2.3in]{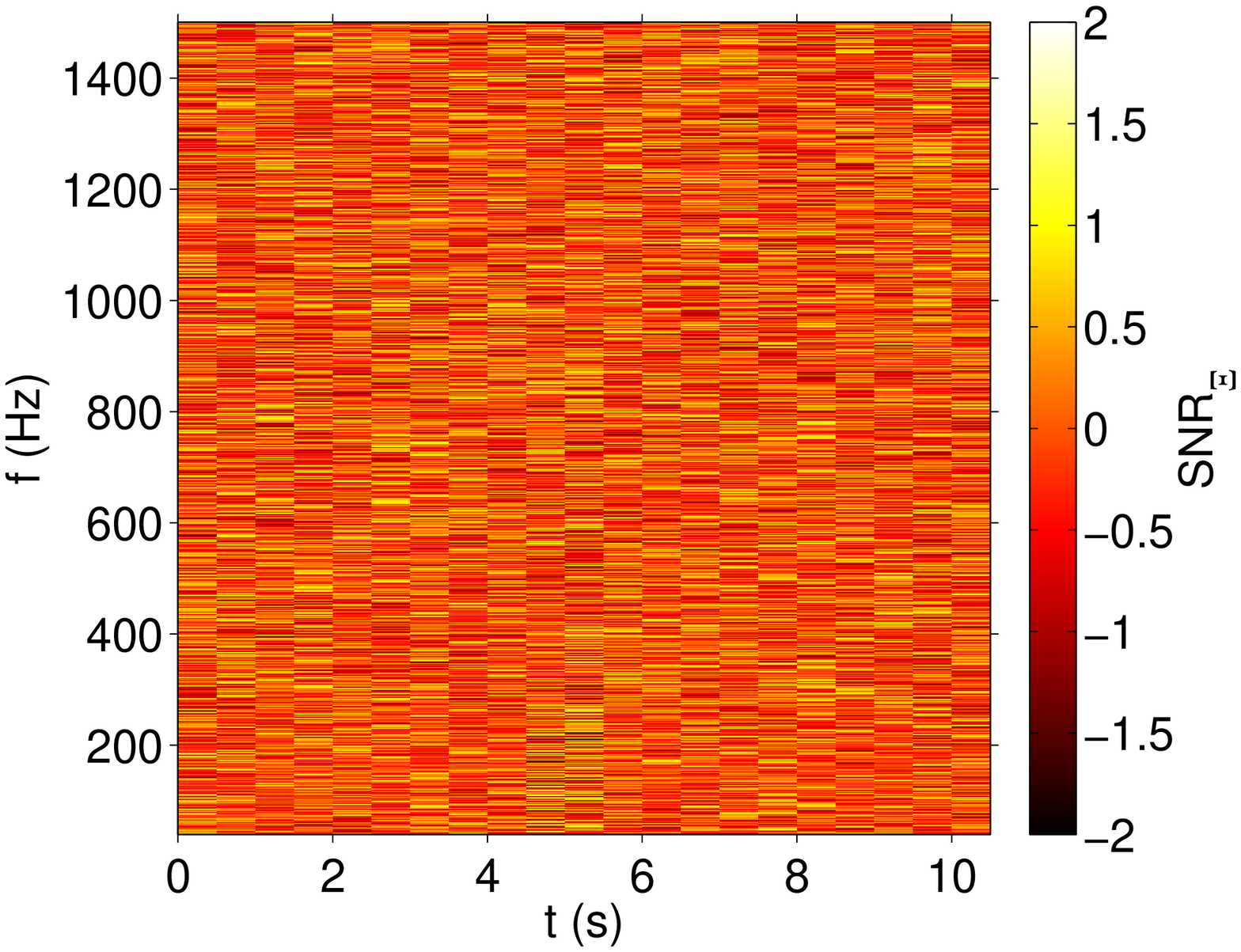}} \\
  \caption{$Ft$-maps of $\text{SNR}$ (left) and $\text{SNR}_\Xi$ (right) for a $\approx{1}{s}$ accretion disk fragmentation waveform injected into stationary noise ($d=\unit[1]{Mpc}$).}
  \label{fig:piro}
\end{figure}

The fragmentation model can be tuned to produce shorter burst-like signals.
Burst-like signals present an extra challenge to the glitch identification algorithm because, like a glitch, the power is typically concentrated in a single ${\cal O}(\unit[1]{s})$-wide $ft$-map column (though we still expect the auto-power between two interferometers to be consistent for a well-constructed filter).
While we are primarily concerned here with {\em long} transients, we use a short $\approx\unit[1]{s}$ fragmentation waveform in Section~\ref{safety} in order to study the performance of the glitch identification flag in this limiting case.
We shall see that, while the flag performs best for long transients, the false dismissal rate is low even for short signals unless the signal is unrealistically loud.
The fragmentation waveforms from~\cite{lucia} are parameterized by the mass of the central black hole $M$, the torus scale height $\eta$, the torus viscosity $\alpha$ and the initial radius $r_0$ (in units of black hole mass).
We use $M=10M_\odot$, $\eta=0.8$, $\alpha=0.1$ and $r_0=200$ to create the $\approx\unit[1]{s}$ waveform here.
$Ft$-maps of this fragmentation waveform are shown in Fig.~\ref{fig:piro}.

\section{Safety}\label{safety}
A critical aspect of any glitch identification algorithm is its \emph{safeness}: the probability that it falsely identifies a segment associated with a GW signal as glitch-like.
To test the safeness of our glitch flag, we apply it to ADI injections in Gaussian simulated noise at dif{\kern0pt}ferent sky locations.
Many long transient signals (including the ADI model considered here) are expected to be elliptically polarized~\cite{stamp}.
In practice, however, it is possible to search for such signals with an unpolarized filter since the two-detector statistic $\hat{Y}(t;f)$ is largely unaf{\kern0pt}fected by polarization details, if the signal is not so long that the polarization degeneracy is resolved by the rotation of the Earth.
In this analysis we use circularly polarized waveforms, a plausible model for many elliptically polarized sources with electromagnetic triggers, which tend to be observed head-on~\cite{kobayashi}.

\begin{figure}[hbtp!]
  \centering
  \includegraphics[height=2.5in]{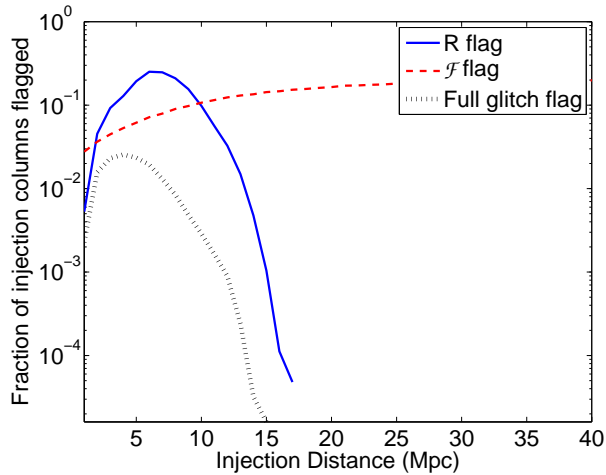}
  \caption{Safety study for simulated accretion disk instability signals.  The $x$-axis is the distance to the source.  For each distance, we average over $312$ directions and $20$ noise realizations.  The $y$-axis is the fraction of segments satisfying the $R$ criteria (solid blue), the $\text{SNR}_\Xi$ criteria (dashed red) and satisfying both in such a way as to be flagged as glitch-like (dotted black).  Note that the fraction of segments flagged as glitch-like decreases at closer distances (corresponding to louder signals) because both detectors exceed the threshold on $R(t)$, which prevents the signal from being flagged as glitch-like (see Eqs.~\ref{eq:flag_params1},\ref{eq:flag_params2}).}
  \label{fig:safety}
\end{figure}

In Fig.~\ref{fig:safety} we present the results of a safety study in which we perform Monte Carlo injections of ADI signals on top of Gaussian simulated noise.
We use $312$ uniformly distributed sky directions with $20$ noise realizations for each direction.
To be flagged as glitch-like, a segment must satisfy our requirements on $R(t)$ and $\text{SNR}_\Xi$ (see Eqs.~\ref{eq:flag_params1},\ref{eq:flag_params2}).
For each injection we record the fraction of segments satisfying the requirements on $R(t)$ alone (blue), $\text{SNR}_\Xi$ alone (red) and segments meeting both criteria and therefore being identified as glitch-like (black).
Note that our ADI signal spans $\approx39$ data segments.
For marginally detectable signals (a $d=\unit[38]{Mpc}$ signal can be recovered with $p=0.1\%$), the fraction of flagged segments is negligible.
For a very loud signal at $d=\unit[5]{Mpc}$ (see the lower-left-hand plot in Fig.~\ref{fig:Xi_glitch}), the fraction of flagged segments becomes $3\%$.
We conclude that for realistic (marginally-detectable signals), the proposed glitch identification flag leads to a acceptably small false dismissal rate.
In order to further reduce the false dismissal rate for very high-$\text{SNR}$ signals, one could design a less aggressive auto-power cut for triggers with extremely high $\text{SNR}_\Xi$, but this is beyond our present scope.

We also consider the case of the short $t\approx\unit[1]{s}$ accretion disk fragmentation signal described in Section~\ref{toy}.
In order to test the glitch rejection algorithm on this short signal, we inject the waveform on top of Monte Carlo noise.
We vary the distance of the injection and perform many trials at each distance, averaging over sky location.
For a very loud $d=\unit[1]{Mpc}$ signal, the false dismissal probability is high: 21\%.
However, we find that false dismissal probability is $<1\%$ for signals at $d>\unit[2.4]{Mpc}$.
While our clustering algorithm is not designed for signals that are vertical $ft$-map columns, we can estimate our sensitivity to short signals by summing all the pixels in the brightest column in order to calculate a total $\text{SNR}$ for that segment~\cite{stamp}.
For signals at $d=\unit[2.4]{Mpc}$, the total $\text{SNR}\approx12$ on average.
For a quasi-normally distributed quantity like total $\text{SNR}$, this corresponds to an extremely small $p$-value.
We conclude that even for very short signals, the false dismissal probability is small for signals with realistic values of $\text{SNR}$, though, unrealistically high values of $\text{SNR}$ have a significant probability of being flagged as glitch-like.

\begin{figure}[hbtp!]
  \centering
  \subfloat[]{\includegraphics[height=2.3in]{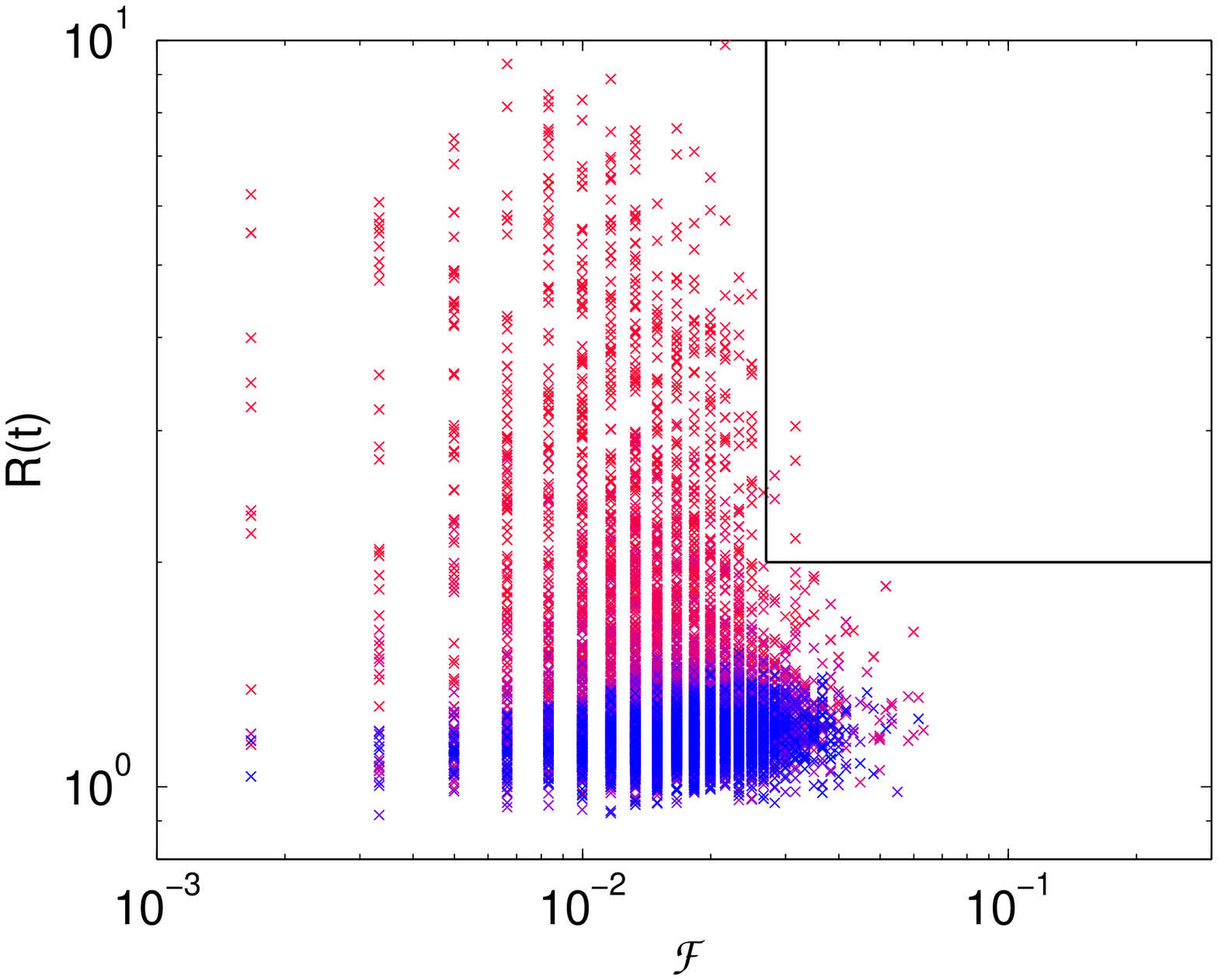}} \hspace{7pt}
  \subfloat[]{\includegraphics[height=2.3in]{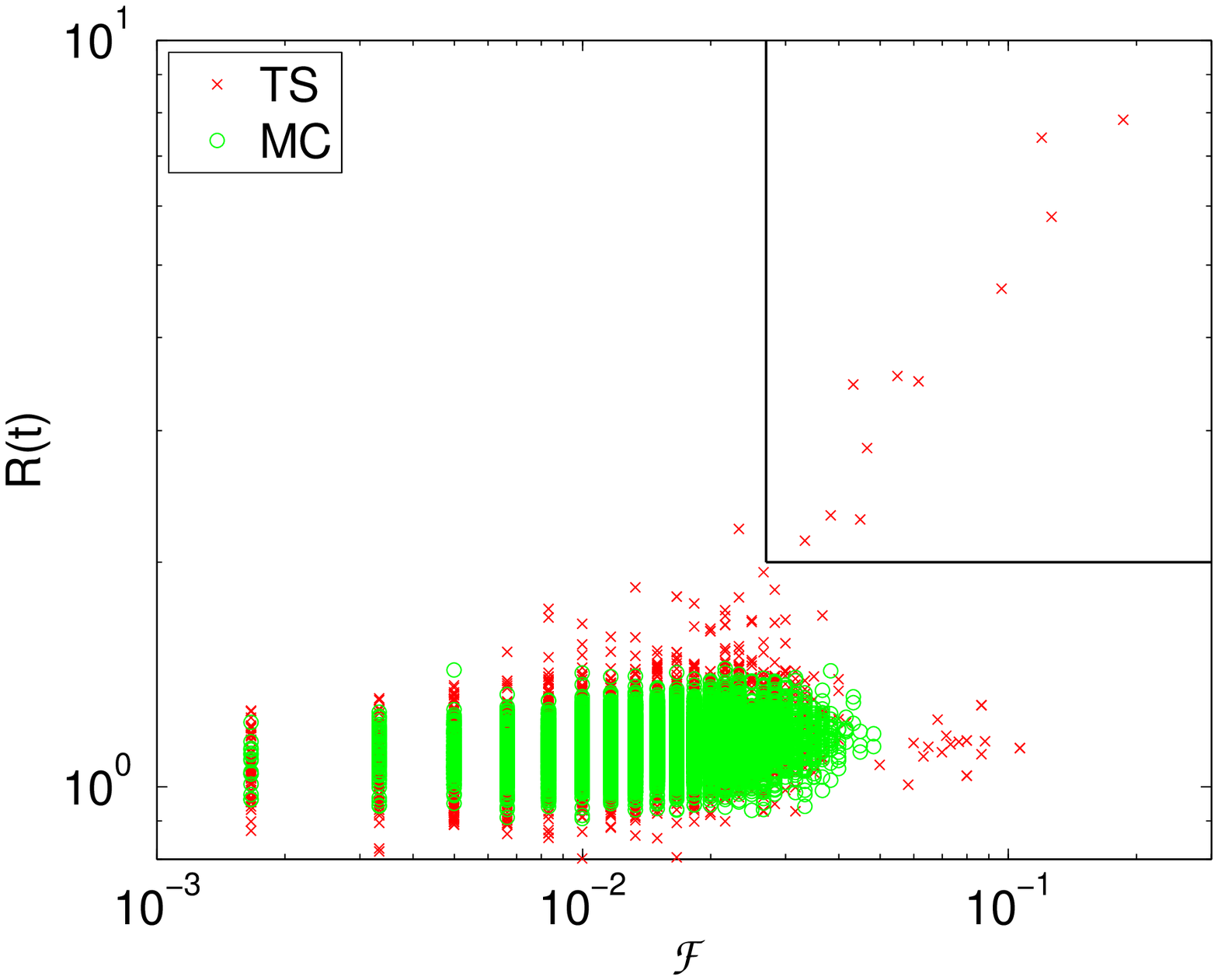}} \\
  \caption{Scatter plots of injected ADI signals (left) and time-shifted data (right red $\times$'s) and Monte Carlo noise (right green $\circ$'s) in the plane of our glitch identification parameters $R(t)$ and ${\cal F}$.  Injection distances range from $\unit[5-40]{Mpc}$ with smaller distances corresponding to redder data markers.  The glitch identification thresholds for each parameter are represented by black lines and points in the upper right quadrant are flagged as glitch-like.   Note that ${\cal F}$ takes on discrete values as discussed below Eq.~\ref{eq:criteria}.}
  \label{fig:scatter}
\end{figure}

Having discussed both the ef{\kern0pt}ficacy of the algorithm flagging glitches as well as its safeness {\em not} flagging segments associated with GW signals, it is interesting to consider the parameter space of the cut.
In Fig.~\ref{fig:scatter}, we show scatter plots of injected ADI signals (left) and noise (right) in the plane of our glitch identification parameters $R(t)$ and ${\cal F}$.
The horizontal and vertical lines indicate the glitch-likely thresholds.
Data markers in the upper right-hand quadrant satisfying both cuts are flagged as glitch-like.
The left-hand plot includes eight dif{\kern0pt}ferent ADI injection distances ranging from $d=\unit[5]{Mpc}$ to $\unit[40]{Mpc}$; redder data markers correspond to smaller distances.
We consider injections from $50$ random directions at each distance, each of which is associated with $20$ time segments, giving a total of $20\times50\times8=8000$ data markers.
The right-hand plot includes $8000$ data markers for S5 LIGO time-shifted data (red $\times$'s) and $8000$ data markers for Monte Carlo Gaussian noise (green $\circ$'s).
Our cut is chosen to exclude the ``glitch tail'' of the red time shift distribution extending up and to the right while preserving most of the injected signals.
Dif{\kern0pt}ferent signal models and dif{\kern0pt}ferent noise environments may require dif{\kern0pt}ferent cuts than the ones presented here.

\section{Comparison with other data-quality flags}\label{DQ}
As noted above, numerous methods have been devised in order to determine when the strain channel is contaminated or corrupted by environmental or subsystem noise (see, e.g.,~\cite{lsc_glitch,smith_glitch,ajith_glitch,stamp_pem,isogai,ballinger,christensen,slutsky}).
A natural question, therefore, is: to what extent does the glitch identification flag developed here provide information complementary to existing data-quality flags?
During the S5 science run, LIGO data quality flags were classified in terms of numbered categories $1-4$~\cite{lsc_glitch,slutsky}.
These four categories describe different levels of severity: Category 1, which includes data that will not be analyzed as it is corrupted or contaminated by known and identified processes; Category 2, where the data is analyzed but various vetoes~\cite{lsc_glitch,smith_glitch,ajith_glitch,stamp_pem,isogai,ballinger,christensen,slutsky} will be applied only in post-processing; Category 3, which are advisory flags used for detection confidence;  and Category 4, which are advisory flags used to exert caution in case of a detection candidate.
Comprehensive descriptions of the S5 data quality flags are fully described elsewhere~\cite{lsc_glitch,slutsky}.

\begin{figure}[hbtp!]
  \centering
  \includegraphics[height=2.5in]{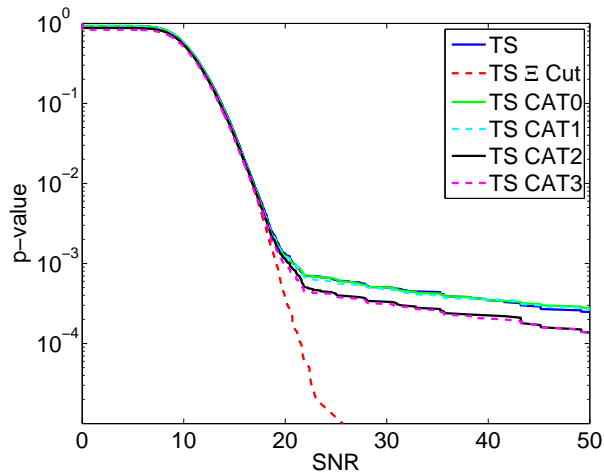}
  \caption{
      A plot of $p$-value vs.\ $\text{SNR}$ for $\unit[1]{s}\times\unit[1]{Hz}$ resolution time-shifted data (TS) with no flags applied (solid blue), with the $\text{SNR}_{\Xi}$-based flag applied (dashed red), and with the data quality flags~\cite{lsc_glitch,slutsky} applied in succession (CAT0 representing no flags applied, CAT1 representing the application of the Category 1 flags, CAT2 representing Category 1 and 2 flags, etc.).
      The data are parsed into $\unit[12]{s}\times\unit[150]{Hz}$ $ft$-maps.
    }
    \label{fig:detchar}
\end{figure}

The numbering is meant to convey the usability of the data, with Category~1 flags representing the most contaminated data.
In Fig.~\ref{fig:detchar} we plot $p$-value vs.\ $\text{SNR}$ for time-shifted data with no flag applied (solid blue), with $\text{SNR}_{\Xi}$-based flag applied (dashed red), and with various data quality flag categories applied in succession (no flags, Category 1 applied, then Categories 1 and 2 applied, etc).
We find that the $\text{SNR}_{\Xi}$-based flag removes a significant number of glitches that are not already identified by category-numbered flags.
It is evident that the two types of flags are complementary---our glitch identification flag finds inconsistencies in autopower between detectors while the category-numbered flags identify and characterize specific instrumental and environmental fluctuations.

\section{Conclusions}\label{conclusions}
There is strong motivation for searches for long unmodeled GW transients, but searches utilizing an excess cross-power statistic~\cite{stamp} must contend with glitches, which hamper sensitivity.
We introduce an auto-power consistency algorithm for identifying glitch-like data segments in searches for long GW transients and we study its behavior in various regimes: well-behaved noise, glitchy noise and potentially detectable GW signals.
We find that it is ef{\kern0pt}fective at identifying glitches with minimal losses in data and live-time, thereby improving sensitivity.
Yet it is safe in the sense that it does not flag GW signals at a high rate.
Finally, we note that the glitch identification algorithm presented here may be useful for searches for short-duration transients.
This is an area of ongoing research.

\ack{
  We thank Peter Kalmus and Rubab Khan, authors of the {\tt BurstCluster} algorithm.
  This work was supported by NSF grants PHY-0854790 and PHY-0758035.
  This is LIGO document \#P1100129.
}

\begin{appendix}
\section{Additional formalism}\label{app}
We consider the general form of a metric perturbation from a point source in the transverse-traceless gauge $h_{ab}(t,\vec{x})$.
It can be written in terms of GW field Fourier coef{\kern0pt}ficients, $\tilde{h}_A(f)$:
\begin{equation}\label{eq:h_ab}
  h_{ab}(t,\vec{x}) = 
  \sum_A \int_{-\infty}^\infty df\, e^A_{ab}(\hat\Omega) \, 
  \tilde{h}_A(f) \, e^{2\pi i f(t+\hat\Omega\cdot \vec{x}/c)} .
\end{equation}
Here $t$ is time, $\vec{x}$ is the position vector, $\{e_{ab}^A\}$ are the GW polarization tensors, $\hat\Omega$ is the direction to the source and $A$ runs over $+$ and $\times$ polarizations.
The dependence of $h_{ab}(t,\vec{x})$ on $\hat\Omega$ is implicit.
The GW strain power between times $t$ and $t+\delta t$ in some frequency band between $f$ and $f+\delta f$ is
\begin{equation}\label{eq:HAAdef}
  H_{AA'}(t;f) = \frac{2}{\cal N}
  \langle \tilde{h}_A(t;f) \tilde{h}_{A'}(t;f) \rangle .
\end{equation}
The factor of two comes from the fact that we consider the single-sided power spectrum and ${\cal N}$ is a normalization factor arising from the use of a discrete Fourier transform.
The semicolon emphasizes that $t$ refers to the beginning a data segment of length $\delta t$ and not to the many sampling times associated with each segment.
Following~\cite{stamp}, we define $H(t;f)$ so as to be invariant under change of polarization basis:
\begin{equation}\label{eq:Hdef}
  H(t;f) = \text{Tr}\left[H_{AA'}(t;f)\right] .
\end{equation}

The metric perturbation in Eq.~\ref{eq:h_ab} induces a strain in detector $I$ given by
\begin{equation}
  \tilde{h}_I(t;f) = \sum_A \tilde{h}_A(t;f,\hat\Omega)
  e^{2\pi i f(t+\hat\Omega\cdot \vec{x}_I/c)}
  F_I^A(t;\hat\Omega) .
\end{equation}
Here $F_I^A(t;\hat\Omega)$ is the antenna factor for detector $I$ (see~\cite{allen-romano}) and $\vec{x}_I$ is its position vector.
The measured strain in detector $I$ is given by the sum of $\tilde{h}_I(t;f)$ with a noise term $\tilde{n}_I(t;f)$:
\begin{equation}
  \tilde{s}_{I}(t;f) = \tilde{h}_I(t;f) + \tilde{n}_I(t;f) .
\end{equation}
We assume that the noise in two interferometers is uncorrelated, which is easily achieved for spatially separated interferometers.

In~\cite{stamp} it was shown that one can construct an unbiased estimator for $H(t;f)$ using the cross-power $\hat{C}_{IJ}(f)$ created from two spatially separated interferometers $I$ and $J$.
This estimator is given by
\begin{equation}\label{eq:y_stat}
 \fl \hat{Y}(t;f) = \text{Re}\left[Q_{IJ}(t;f,\hat{\Omega},\vec\alpha)\,\hat{C}_{IJ}(t;f) \right] =
  \frac{2}{\cal N} \text{Re}\left[
    Q_{IJ}(t;f,\hat{\Omega},\vec\alpha) \, \tilde{s}_I^\star(t;f) \tilde{s}_J(t;f)
    \right] .
\end{equation}
Here $Q_{IJ}(t;f,\hat\Omega,\vec\alpha)$ is a filter function which takes into account the phase delay from the spatial separation of the interferometers as well as the detection ef{\kern0pt}ficiency of interferometers $I$ and $J$.
It also depends on $\vec\alpha$, which is a set of parameters that characterizes the expected form of $H_{AA'}(f)$ such as the polarization of the source.
We can write the filter function as:
\begin{eqnarray}
  Q_{IJ}(t;f,\hat\Omega,\vec\alpha) & = &
  \frac{1}{\epsilon_{IJ}(t;\hat\Omega,\vec\alpha)}
  e^{2\pi i f \hat\Omega\cdot\Delta\vec{x}_{IJ}/c}
\end{eqnarray}
Here $\Delta\vec{x}_{IJ}\equiv\vec{x}_I-\vec{x}_J$ is the dif{\kern0pt}ference in position vectors for detectors $I$ and $J$.
$\epsilon_{IJ}(t;\hat\Omega,\vec\alpha)$ is the ``pair ef{\kern0pt}ficiency,'' which is defined in terms of the expectation value of interferometer cross- and auto-powers:
\begin{eqnarray}
  \langle \hat{C}_{IJ}(t;f) \rangle &
  \equiv & \epsilon_{IJ}(t;\hat\Omega,\vec\alpha) H(t;f)
  e^{-2\pi i f \hat\Omega\cdot\Delta\vec{x}_{IJ}/c} \\
  \langle\hat{P}_I(t;f)\rangle &
  \equiv & \epsilon_{II}(t;\hat\Omega,\vec\alpha) H(t;f) + N_I(t;f)
  \label{eq:autopower_def}
\end{eqnarray}
where $N_I(t;f)\equiv(2/{\cal N})|\tilde{n}_I(t;f)|^2$ and $H(t;f)$ is defined in Eq.~\ref{eq:Hdef}.
The pair ef{\kern0pt}ficiency for an unpolarized source is:
\begin{equation}
  \epsilon_{IJ}(t;\hat\Omega,\text{unpolarized}) =
  \frac{1}{2} \sum_A F_I^A(t;\hat\Omega) F_J^A(t;\hat\Omega) .
\end{equation}
Hereafter we abbreviate $\epsilon_{IJ}(t;\hat\Omega,\vec\alpha)$ as simply $\epsilon_{IJ}$.

Through our definition of $Q_{IJ}(t;f,\hat\Omega,\vec\alpha)$, we implicitly assume that the direction of the source $\hat\Omega$ is known.
In order to estimate how well we must know $\hat\Omega$, we consider how large the error in $\hat\Omega$ (denoted $\delta\theta$) must be before we lose too much signal power.
If we demand that we measure a fraction of at least $R$ of the total possible power, then we can tolerate angular errors $\delta\theta$ of
\begin{equation}
  \delta\theta \lesssim \cos^{-1}(R)
  \left(\frac{c}{2\pi f\left|\Delta\vec{x}_{IJ}\right|}\right) .
\end{equation}
For the Hanford-Livingston pair, this implies that we can tolerate angular errors of $\delta\theta\lesssim0.8^\circ$ up to $\unit[500]{Hz}$ with $R=90\%$.
For comparison, we note that the Swift experiment has an angular resolution of $\approx0.25^\circ$~\cite{swift}.
For the remainder of the paper, we consider a single search direction.
For triggers with large error regions on the sky, one can iterate over a grid of points inside a search cone, but this is a trivial extra step.

Since by assumption the noise in detectors $I$ and $J$ is uncorrelated, it follows that
\begin{equation}
  \langle \hat{Y}(t;f) \rangle = H(t;f) .
\end{equation}
An estimator for the variance of $\hat{Y}(t;f)$ is given by~\cite{stamp}:
\begin{equation}
  \hat\sigma(t;f)^2 = \frac{1}{2} \left|Q_{IJ}(t;f,\hat\Omega,\vec{\alpha})
  \right|^2 \hat{P}_I'(t;f) \hat{P}_J'(t;f) .
\end{equation}
Here $\hat{P}_I'(t;f)$ and $\hat{P}_J'(t;f)$ are the auto-powers measured in detectors $I$ and $J$, respectively.
The prime denotes that they are calculated using $2n$ neighboring segments in order to obtain an estimate of the noise associated with the segment beginning at $t$:
\begin{equation}\label{eq:pprime}
  \hat{P}_I'(t_0;f) \equiv
  \frac{1}{2n}\left[\sum_{t=t_0-n\delta t}^{t=t_0+n\delta t} \hat{P}_I(t;f) \right]-
  \frac{1}{2n}  \hat{P}_I(t_0;f) .
\end{equation}

\end{appendix}

\bibliography{glitch}

\end{document}